\documentclass[a4paper,fleqn]{cas-dc}

\usepackage{algorithm}
\usepackage{algpseudocode}
\usepackage[numbers]{natbib}

\usepackage{amsmath,amssymb,bm}
\usepackage{physics}   %
\usepackage{siunitx}   %

\usepackage{tikz}
\usepackage{amsmath}
\usepackage{bm}
\usetikzlibrary{arrows.meta, calc, patterns}

\usepackage{float}
\usepackage{placeins}
\usepackage{booktabs}
\usepackage{array}
\usepackage{xcolor}

\newcommand{\gdot}{\dot\gamma}
\newcommand{\deq}{d_{\mathrm{eq}}}

\newcommand{\Rey}{\mathrm{Re}}

\newcommand{\bF}{\mathbf{F}}
\newcommand{\bT}{\mathbf{T}}

\begin{document}
\let\WriteBookmarks\relax
\def\floatpagepagefraction{1}
\def\textpagefraction{.001}

\shorttitle{Drag, lift and torque closures for oblate particles in shear flow}
\shortauthors{Zucchelli F. et~al.}

\title[mode=title]{Drag, lift and torque correlations for monodispersed suspensions of oblate particles in linear
shear flow at low Reynolds numbers}

\author[1]{Francesca Zucchelli}
\cormark[1]
\ead{francesca.zucchelli@unige.ch}
\author[1]{Bastien Chopard}
\author[1]{Jonas Latt}

\affiliation[1]{organization={Computer Science Department, University of Geneva},
                addressline={1227 Carouge},
                country={Switzerland}}

\cortext[1]{Corresponding author}

\begin{abstract}
The fidelity of unresolved CFD--DEM simulations of particle-laden flows is subordinate to the closures describing the hydrodynamic loads on each particle, yet no closure is currently available for oblate particles in crowded suspensions under linear shear flow at low Reynolds numbers. In this work, particle-resolved lattice Boltzmann simulations were performed on fixed, isotropic random suspensions of oblate ellipsoids of aspect ratio $0.25$, with solid volume fractions from $0.2\%$ to $15\%$ and slip Reynolds numbers $0.1\le\Rey\le1$, complemented by isolated oblate simulations. Moving the isolated particle across the domain height allowed the effects of slip and shear to be separated: the force was found to depend on the local slip alone, whereas slip inertia gives rise to a pitching torque agreeing with exact theory. In suspension, the drag retains the $\cos^2\theta$ dependence on the incidence angle of the isolated oblate,  but crowding raises the axial resistance more strongly than the transverse one: the face-on to edge-on drag ratio grows from $1.27$ (isolated oblate) to $1.66$ at $\phi=15\%$, an orientation-dependent drag enhancement that conventional spherical voidage corrections fail to capture, even at low solid volume fractions. The lift follows directly from the drag anisotropy without any fitted constant, while crowding affects the viscous Jeffery torque and the inertial pitching torque differently. The resulting closure is anchored on the exact Stokes resistances, to which it reduces as $\Rey\to0$ and $\phi\to0$, requires eleven fitted constants, reproduces the mean drag within $1.7\%$ for $\phi\ge2\%$ and is ready to be used in CFD--DEM simulations of particle-laden flows.
\end{abstract}


\maketitle

\section{Introduction}\label{sec:intro}
Particle suspensions are everywhere around us, silently part of every aspect of life from the most fundamental, like blood, to the most seemingly trivial, like snow globes. As they are at the core of many natural and man made processes, accurately resolving particle laden flows is fundamental to the success of industrial and biomedical practices. The handling of such large numbers of particles is generally entrusted to Eulerian-Lagrangian simulations, otherwise known as computational fluid dynamics-discrete element method simulations (CFD-DEM), whose fidelity is subordinate to the quality of the closures designed to describe the hydrodynamic forces and torque on each particle \cite{zhu2007, zhong2016}. Historically, single spheres in flow have been the foundation of suspension studies, despite research on non-spherical particles being almost concurrent. This predominance is largely attributed to the geometrical simplicity of the sphere, whose shape requires only one characteristic dimension for its description, making spherical approximation far more manageable from a modelling perspective. Real particles however, are considerably less perfect in shape and approximating their hydrodynamic force and torque to the ones of a single sphere fully disregards the impact of shape and orientation, as well as crowding, on suspension dynamics. For instance, in blood flow modeling, an accurate representation of particle geometry can have a substantial impact on platelet margination rates, influencing the accurate prediction of transient phenomena in blood flow \cite{zucchelli2026}. Consequently, developing closure models tailored to the targeted particle suspension, which account for accurate particle shape, orientation and suspension microstructure, is fundamental to improve the predictive capabilities of CFD-DEM simulations across application-specific conditions.  

The drag on an isolated sphere at vanishing Reynolds number was first derived by Stokes \cite{stokes1851}, then corrected to account for fluid inertia by Oseen's derivation \cite{oseen1910} and finally extended to larger Reynolds numbers up to  $Re \approx 800$ by the empirical correlation of Schiller-Naumann, which is currently the working standard \cite{schiller1933}. Still in the Stokes limit, investigations on non-spherical particles started with Oberbeck's derivation of the force on a translating ellipsoid \cite{oberbeck1876}, later given for spheroids in closed form by Happel\&Brenner \cite{happel1983}, and Edwardes' torque on a rotating ellipsoid \cite{edwardes1892}, while Jeffery studied the torque of an ellipsoid in linear shear flow \cite{jeffery1922}. Two additional results bear important consequences for the study of particles in shear: a linear shear flow acts on the force only through the slip at the particle's center as theorized on spheres by Faxén and generalized to ellipsoids by Brenner\cite{faxen1922, brenner1964, kim1991} and a centrosymmetric particle feels no force from the stretching part of shear \cite{happel1983, brenner1964}. Hence, in Stokes regime, the force on a spheroid
in linear shear is the same it would be in a uniform flow equal to the slip at the particle center, whereas shear rate enters only through the torque, exactly as the Jeffery torque.

When moving out of the Stokes limit, at small but finite Reynolds numbers fluid inertia must be accounted for, therefore correcting the above mentioned results. The two sources of inertia, namely the slip velocity and the velocity gradient of a shear flow, have traditionally been treated separately. For the slip, building on Oseen's theory, Brenner showed that the drag of an arbitrary particle translating along one of its principal axes increases by a factor proportional to its Stokes drag \cite{brenner1961}. Cox and Khayat have demonstrated on slender and spherical bodies that slip acts on the orientation of the particles, as it produces a torque that turns a translating spheroid broadside-on \cite{cox1965, khayat1989}. The same theory was later generalized for any aspect ratio by Dabade et al. \cite{dabade2015}. On the other hand, Saffman derived the lift produced on a sphere by shear, concluding that it has the same direction as the velocity gradient and is only valid when shear dominates inertia in the far field \cite{saffman1965}. The extension to particles of arbitrary shape was carried out by Harper and Chang \cite{harper1968}. Finally, shear inertia also acts on orientation, making a free spheroid drift away from the Jeffery orbits towards preferred rotation states \cite{subramanian2005,subramanian2006,einarsson2015,rosen2015,dabade2016}.

The exact theory recalled so far however, is confined to small or vanishing Reynolds numbers and mostly idealized shapes, greatly limiting its applicability to practical scenarios. To enable Eulerian-Lagrangian simulations of any real life application, closure relations for hydrodynamic forces and torque covering a wider range of regimes and geometries are needed, thus leading to the development of empirical correlations, followed more recently by relations derived from particle-resolved numerical simulations.

In uniform flow, the earliest proposed drag relations were authored by Haider\&Levenspiel and Ganser who described shape through sphericity and equivalent spherical diameter, leading to the assignment of the same drag force to particles of different shape, solely because they shared the same geometrical parameters \cite{haider1989, ganser1993}. Later, Hölzer and Sommerfeld used the crosswise and lenghtwise sphericity to establish a new drag correlation that included particles orientation through the area projected on the flow \cite{holzer2008}. Orientation started to be explicitly accounted for by particle-resolved simulations in uniform flow as the ones performed by Zastawny et al.\ who correlated drag, lift, pitching and rotational torque for four shapes (two prolate ellipsoids, a disc and a fibre) at $0.1\le Re\le300$ \cite{zastawny2012}. Instead, Ouchene et al.\ treated prolate spheroids with aspect ratios up to 32 at $Re\le240$ \cite{ouchene2016}, while Sanjeevi et al.\ focused on a prolate, an oblate spheroid and a spherocylinder up to $Re=2000$ \cite{sanjeevi2018}, and Fröhlich et al.\ studied prolate spheroids at $1\le Re\le100$ \cite{frohlich2020}. It is worth pointing out that in Stokes flow, the resistance tensor of a spheroid fixes both the values of the drag for flow along and across the particle symmetry axis $C_{D,0}$ and $C_{D,90}$, as well as its dependence on the incidence angle $\theta$ between flow and symmetry axis as per Happel\&Brenner \cite{happel1983}. The link of the above mentioned correlations to these exact results varies greatly. Drag and lift by Fröhlich et al. are built on exact Stokes values, while in the study of Ouchene et al.\ only drag is, whereas the relations of Zastawny et al.\ and Sanjeevi et al.\ are compared with, but not built on, them. Moreover, the $\sin^2\theta$ scaling of the drag was found to persist far beyond Stokes flow, up to $Re=2000$ as demonstrated by Sanjeevi et al.\ \cite{sanjeevi2018, sanjeevi2017} for prolate spheroids and mildly oblate ones. The lift scaling however is more fragile: at finite $\Rey$ the maximum of the lift has been found to move away from $\theta=45^\circ$ for prolate particles \cite{ouchene2016,frohlich2020}, while for oblate spheroids the $\sin\theta\cos\theta$ form is only valid at low Reynolds numbers so the resistance tensor cannot be relied upon to carry the lift above this limit \cite{sanjeevi2018}. The torque, by contrast, has no Stokes anchor: in a uniform stream the Stokes torque on a fixed spheroid vanishes identically, so all these torque correlations are purely inertial, and the Jeffery torque is absent by construction.

In shear flow, single particle hydrodynamic correlations are fewer and explore both unbounded flow and wall closeness. Kurose\&Komori applied a finite difference scheme to treat the case of a sphere in unbounded linear shear, separating the dimensionless shear rate $\tilde G=d\dot\gamma/U$ from the Reynolds number $Re = d_{eq}U/\nu$ \cite{kurose1999}, while Hölzer and Sommerfeld computed the loads on non-spherical particles by the lattice Boltzmann method \cite{holzer2009}. For prolate rods with aspect ratios between 2.5 and 10, Chéron et al.\ gave drag, lift and torque correlations for $0.1\le Re\le300$, again defining $\tilde G$ as a separate variable. They restricted particle orientation to the shear plane and closed the viscous regime by exact theory \cite{cheron2024a}. A shear flow closure anchored to Stokes flow, with slip and shear separated, thus exists but it is limited to isolated prolate rods and accounts for shear as an additive correction of a uniform flow loads.

Near a wall the picture changes as the symmetry of the particle-fluid system gets broken by the closeness to a surface, and hydrodynamic forces and torque acquire additional components with respect to the unbounded case. Classical results are the ones by O'Neill, who obtained the exact creeping flow solution for a sphere touching a wall \cite{oneill1968}, and Zeng et al.\ who computed the forces on a sphere close to a wall at finite $\Rey$ \cite{zeng2009}. A similar study involving rods was carried out by Chéron et al.\ \cite{cheron2024b}, while a series of recent works have addressed oblate spheroids from different standpoints: attached to the wall at finite $\Rey$ \cite{cao2025}, in contact with the wall and rotated about its normal for both forces \cite{desouza2024, desouzaErratum2026} and torque \cite{desouza2025}, at varying wall distances and rotated about the spanwise axis \cite{zhang2026}. In all of these oblate studies, $\Rey$ is defined by the undisturbed velocity at the particle centroid, closely tying together slip velocity and shear rate which vary in fixed proportion at a given wall gap. Therefore, when a particle moves away from the wall and the hydrodynamic loads change, this slip-shear coupling leads to the inability of attributing the responsibility of such change to one or the other player: the change in shear to slip ratio or the wall moving away.

Finally, a common trait to most of these single particle studies, both in uniform and shear flow, is the combination in the transverse force component of three mechanisms of different origin: the orientation-induced force of the Stokes resistance tensor, which is not a lift in the inertial sense \cite{happel1983,kim1991}, the shear-induced inertial lift and the wall-induced force, are merged into a single fitted lift coefficient \cite{zastawny2012,ouchene2016,sanjeevi2018,cao2022}. 

\begin{table*}[t]
  \caption{Literature overview ($\lambda$: particle aspect ratio. $\phi$: volume fraction. $F, T, D, L$: force, torque, drag and lift respectively).}\label{tab:litmap}
  \small
  \begin{tabular}{@{}p{0.2\textwidth}p{0.1\textwidth}p{0.1\textwidth}p{0.09\textwidth}p{0.1\textwidth}p{0.1\textwidth}p{0.08\textwidth}p{0.08\textwidth}@{}}
  \toprule
  Study & Shape, $\lambda$ & Flow & $\Rey$ & $\phi$ & Loads & Stokes anchor & Shear \\
  \midrule
  \citet{jeffery1922,brenner1964,kim1991} & ellipsoid & linear & $0$ & $0$ & $\bF,\bT$ & exact & yes \\
  \citet{dabade2015} & spheroid, any $\lambda$ & uniform & $O(\Rey)$ & $0$ & inertial $T$ & exact & no \\
  \citet{jiang2021} & spheroid $1/6$--$6$ & uniform & $0.3$--$50$ & $0$ & inertial $T$ & --- & no \\
  \citet{haider1989,ganser1993,holzer2008} & general & uniform & broad & $0$ & $D$ & partial & no \\
  \citet{zastawny2012} & prolates, disc, fibre & uniform & $0.1$--$300$ & $0$ & $D,L,T$ & no & no \\
  \citet{ouchene2016} & prolate $1$--$32$ & uniform & $\le240$ & $0$ & $D,L,T$ & yes ($D$) & no \\ 
  \citet{sanjeevi2018} & prolate $5{:}2$, oblate $2{:}5$, spherocyl.\ $4$ & uniform & $0.1$--$2000$ & $0$ & $D,L,T$ & partial & no \\
  \citet{frohlich2020} & prolate, $1$--$8$ & uniform & $1$--$100$ & $0$ & $D,L,T$ & yes ($D,L$) & no \\
  \citet{cao2020,cao2021} & prolate, $2.5$--$10$ & uniform & $10$--$200$ & $0.1$--$0.3$ & $D$ & no & no \\
  \citet{cao2022} & prolate, $2.5$--$10$ & uniform & $10$--$200$ & $0.1$--$0.3$ & $L,T$ & no & no \\
  \citet{sanjeevi2020} & spherocyl.\ $4$ & uniform & $0.1$--$1000$ & $0.1$--$0.5$ & $D,L,T_P$ & no & no \\
  \citet{li2019prolate,li2021oblate} & prolate $1.25$--$5$, oblate $0.2$--$0.8$ & uniform & $<0.1$ & $0.1$--$0.4$ & $D, F_y, F_z$ & prefactors & no \\
  \citet{kurose1999} & sphere & shear & $1$--$500$ & $0$ & $D,L$ & no & $\tilde G$ \\
  \citet{cheron2024a} & prolate rods, $\alpha{=}2.5$--$10$ & shear & $0.1$--$300$ & $0$ & $D,L,T$ & yes & $\tilde G$, in-plane $\theta$ \\
  \citet{cheron2024b} & prolate rods, $\alpha{=}2.5$--$10$ & shear & $0.1$--$200$ & $0$ & $D,L,T$ & partial & wall distance \\
  \citet{desouza2024} & oblate $\lambda{=}0.2$--$1$ & shear & $0.1$--$7$ & $0$ & $D,L$, drift & no & wall contact \\
  \citet{desouza2025} & oblate $\lambda{=}0.2$--$1$ & shear & $0.1$--$7$ & $0$ & $T$ & no & wall contact \\
  \citet{cao2025} & oblate $\lambda{=}1.5,2,4$ & shear & $1$--$200$ & $0$ & $D,L,T$ & no & $\gdot\propto$ slip \\
  \citet{zhang2026} & oblate $\lambda{=}1.5,2,4$ & shear & $1$--$200$ & $0$ & $D,L,T$ & no & wall distance \\
  \citet{derksen2011} & spheres, $1,2,3$ layers & shear & $\le0.05$ & dense layers & $D,L$ & --- & wall contact \\
  \citet{cao2026} & spheres, layers & shear & $1$, $10$, $100$ & $0.065$--$0.40$ & $D,L$ & partial & wall contact \\
  \textbf{Present work} & oblate $\lambda{=}0.25$ & shear & $0.1$--$1$ & $0$--$0.15$ & $D,L_y,L_z,\bT$ & exact & $\gdot$ and slip separated \\
  \bottomrule
  \end{tabular}
  \end{table*}

When it comes to suspensions, one must factor in the presence of multiple particles which hydrodynamically influence each other, so the closures require accounting for the local solid volume fraction $\phi$. However, their simulation is oftentimes approached by combining the single particle closures discussed so far, with a voidage function such as those of Di Felice \cite{difelice1994}, Richardson\&Zaki \cite{richardson1954} or Wen\&Yu \cite{wen1966}. The widely used law by Di Felice for instance, was calibrated at low volume fractions on sedimenting spheres for which the drag is linear in $\phi$, making it particularly ill suited for the description of weakly crowded fixed beds. Indeed, in a fixed bed the disturbance that a particle creates in the flow is screened by its neighbors and as a consequence, at small volume fractions, the drag grows with the square root of $\phi$ rather than linearly. A voidage function instead is linear in $\phi$ at small $\phi$ \cite{brinkman1947, kimrussel1985}. Furthermore, these voidage functions being built on assemblies of spheres, they are unable to account for different particle orientations leading to the under prediction of the loads in the case of suspensions of ellipsoids, as evidenced especially at finite $\Rey$ by the work of He et al. \cite{HeTafti2017}. Again, on the common practice of combining an isolated-particle law with a sphere voidage function, the approach by Sanjeevi et al.\ works up to $\phi=30\%$ for spherocylinders and fails at $50\%$ \cite{sanjeevi2020}, whereas for oblate ellipsoids Li et al.\ under-predict the drag already at $\phi=10\%$, the more so the flatter the particle \cite{li2021oblate}.

Assemblies of non spherical particles gained attention only recently and the available particle resolved closures have all been obtained in uniform flow and at $\phi\ge10\%$. Cao et al.\ focused on prolates and corrected existing isolated particle laws for the drag \cite{cao2020,cao2021}, the lift and torque \cite{cao2022} with volume fraction and orientation-dependent modifiers. Sanjeevi and Padding simulated assemblies of spherocylinders and confirmed that the $\sin^2\theta$ scaling of the drag survives in the suspension up to $\phi= 40\%$, independently of the mutual orientation of the particles. Also, they found that the lift is approximated by the drag difference with a mean deviation of about $12\%$ and at $Re=0.1$ the ratio of the drags across and along the axis stays constant as $\phi$ grows. Both oblate and prolate random arrays were simulated by Li et al.\ at $Re<0.1$ and $\phi$ between $10\%$ and $40\%$ \cite{li2019prolate,li2021oblate}. They correlated the ensemble drag of the whole array along and across its preferred direction and obtained the lateral force by Stokes superposition, without per-particle forces or torque, showing that in aligned oblate arrays the contrast between face-on and edge-on drag grows with $\phi$. Correlations for the torque in a particle suspension are scarce and built in uniform flow, therefore provide only an inertial pitching torque, while the Jeffery torque is naturally absent \cite{sanjeevi2020,cao2022}. Finally, the only examples of studies investigating crowding in shear flow are, to the best of our knowledge, limited to spheres: at low $\Rey$ in random mono and multi layers \cite{derksen2011} and at finite $\Rey$ in a single and double layer where an isolated sphere law is adjusted by fitted drag and lift factors \cite{cao2026}.\\
For additional details on the discussed works, we refer the readers to Table \ref{tab:litmap}.

To summarize, the available literature highlights the following: slip and shear inertia are treated separately by exact theory, single-particle correlations for oblate particles in unbounded shear, with a fully three-dimensional orientation in the range $0.1\le Re\le1$ are currently missing, and existing closures for suspensions of non-spherical particles start at $\phi=10\%$, are formulated in uniform flow only and keep lift separated from drag. 

In this work we propose a closure for the drag, lift and torque acting on oblate particles in suspensions of increasing crowding under linear shear flow at Stokes regime. The motivation of this study stems from the need to simulate large-scale blood flows such as those in stenosed vessels, that are made computationally challenging by the high number of particles involved. For this reason, unresolved CFD-DEM simulations are the natural approach, provided that suitable closures for blood cells are available. Platelets can be reasonably approximated by oblate-like particles, which travel in sheared flows at low Reynolds numbers and are constantly surrounded by neighboring cells. This is a combination that none of the works discussed above addresses as a whole. We performed particle-resolved lattice Boltzmann simulations of oblate spheroids of aspect ratio $0.25$ in random suspensions with volume fractions from the dilute limit up to $15\%$, complemented by simulations of the isolated particle that fix the dilute limit and the inertial torque. The closure was built on the exact Stokes theory and the simulations supplied the corrections due to inertia and crowding, that would otherwise be unattainable. Moving the isolated particle across the channel height, away from the walls, allowed us to separate the effects of slip and shear, which near-wall studies cannot disentangle. The resulting closure shows that neighboring particles act on the two resistances so that the lift on a particle follows directly from its drag and that crowding affects the viscous and the inertial torques differently, providing a single model anchored on theory that is ready to be used in CFD-DEM simulations of platelet-laden flows.

The article is organized as follows: first the numerical method in use is thoroughly described together with details on how the suspensions were generated, then validations of the method and results concerning the simulation of oblate particles in linear shear flow are presented and discussed, finally the derived closure is provided followed by concluding remarks.

\section{Numerical method}\label{sec:method}
Mono dispersed suspensions of fixed oblate ellipsoids in linear shear flow were modeled via particle resolved simulations by the Lattice Boltzmann Method (LBM) to extract the hydrodynamic loads exerted by the fluid on each particle and build the closure model at the core of this work. LBM is quite an established approach when it comes to this class of problems, for isolated particles \cite{holzer2009, sanjeevi2017,sanjeevi2018} and especially for assemblies \cite{derksen2011,sanjeevi2020,li2019prolate,li2021oblate}, as it is particularly suited to being massively parallelized, allowing for the resolution of several thousand particles in suspension in a single domain. The simulations were performed with a modified version of LEDDS, our own GPU accelerated Lattice Boltzmann-Discrete Element Method  framework which has been extensively validated for flow past particles \cite{rambosson2026}. The next subsections will be dedicated to describing the fluid solver, illustrating the fluid-particle coupling and simulation setup.  

\subsection{Lattice Boltzmann method} \label{sec:lbm}
Fluid flow was simulated using the D3Q19 TRT lattice Boltzmann method \cite{ginzburg2008}, for which the evolution of velocity distribution function $f_i(\mathbf{x},t)$ with $i\in\left\{1, ..., 19\right\}$, representing the number of fictitious particles with a given velocity $\mathbf{c}_i$ at point $(\mathbf{x},t)$, is computed on a uniform grid of spacing $\Delta x$ as
\begin{equation}
f_i(\mathbf{x}+\mathbf{c}_i\Delta t,\,t+\Delta t)=\left[f_i+\Omega_i\right](\mathbf{x},t),
\label{eq:lbe}
\end{equation}
where $\Omega_i$ is the collision model in its general form, that in case of TRT reads as
\begin{equation}
\Omega_i = - s_\nu\left(f_i^+ - f_i^{\mathrm{eq}+}\right)
- s^-\left(f_i^- - f_i^{\mathrm{eq}-}\right),
\label{eq:trt}
\end{equation}
with $f_i^{\pm} = 0.5(f_i \pm f_{\bar{i}})$ and $f_i^{\mathrm{eq}\pm} = 0.5(f_i^{\mathrm{eq}} \pm f_{\bar{i}}^{\mathrm{eq}})$ the symmetric and antisymmetric parts of the distribution and equilibrium functions and $\bar{i}$ denoting the direction opposite to $i$ ($\mathbf{c}_{\bar{i}}=-\mathbf{c}_i$). The symmetric part carries the even moments and the antisymmetric part the odd moments. The even moments relax with rate $s_\nu=1/\tau$, which sets the kinematic viscosity $\nu=c_s^2\left(\tau-\tfrac12\right)\Delta t$ with $c_s^2=\Delta x^2/(3\Delta t^2)$, while the odd moments relax with rate $s^-$ chosen through the so-called magic parameter
\begin{equation}
\Lambda=\left(\frac{1}{s_\nu}-\frac12\right)\left(\frac{1}{s^-}-\frac12\right)=\frac{3}{16},
\label{eq:magic}
\end{equation}
for which the location of a bounce-back wall becomes independent of the viscosity \cite{ginzburg2008,kruger2017}. 

The boundary condition on the particle surfaces needed to accurately follow the curved geometry of oblate ellipsoids, hence the linearly interpolated bounce-back scheme of Bouzidi et al.\ \cite{bouzidi2001} was chosen, instead of a more traditional staircase bounce back implementation, to impose a no-slip condition similarly to Hölzer\&Sommerfeld \cite{holzer2009} and Sanjeevi et al.\ \cite{sanjeevi2018, sanjeevi2020}.

The flow was driven by a body force which sustains a linear velocity profile $\mathbf{U}=\dot\gamma\,y$ over the whole domain height. This approach was preferred over the more conventional wall-driven shear, since the fixed particles in suspension screen the flow and would absorb the momentum diffusing from the sliding walls, leaving the core of the domain essentially at rest. The force acts on the mean flow of each plane normal to the gradient direction, so that at every time step, the streamwise velocity is averaged over the $N_f(y)$ fluid nodes of a plane at height $y$ as follows:
\begin{equation}
\langle u_x\rangle(y,t)=\frac{1}{N_f(y)}\sum_{\mathbf{x}\in y}u_x(\mathbf{x},t),
\label{eq:planemean}
\end{equation}
Every fluid node of that plane receives the same streamwise momentum increment, proportional to the departure of this mean from the target profile $U_0(y)=\dot\gamma_0\,y$,
\begin{equation}
\delta j_x(y,t)=\alpha\,\rho_0\left[U_0(y)-\langle u_x\rangle(y,t)\right],
\label{eq:forcing}
\end{equation}
which is added to the populations entering the collision step as
\begin{equation}
f_i(\mathbf{x},t)\;\leftarrow\;f_i(\mathbf{x},t)+\frac{w_i\,c_{ix}}{c_s^2}\,\delta j_x(y,t),
\label{eq:forcing_pop}
\end{equation}
with $w_i$ the lattice weights, $\rho_0$ the reference density and $\alpha=0.1$ the gain of the forcing. Since $\sum_iw_ic_{ix}=0$ and $\sum_iw_ic_{ix}^2=c_s^2$, Eq.~\eqref{eq:forcing_pop} injects exactly the momentum $\delta j_x$ and no mass, i.e.\ it is equivalent to a body force $\delta j_x/\Delta t$ uniform within each plane. Because the increment depends on the plane mean only, it is the same at every node of a plane and does not act on the local disturbance flow around the particles, so that it does not bias the loads measured on them. The target shear rate $\dot\gamma_0$ sets the nominal profile and the gain $\alpha<1$ relaxes the plane mean towards it gradually, avoiding the overshoot that a prescription of the mean at every step would produce while the flow develops. Even so, fixed platelets resist the flow, causing a deficit on the velocity profile that grows with the total drag of the suspension. Every quantity was therefore referred to the achieved profile rather than the nominal one. It is worth noting that the achieved velocity is exactly the fluid velocity that an unresolved CFD-DEM solver supplies to a particle, so that the closure is built on the same variable it will be fed with.

\subsection{Generation of particle suspension}\label{sec:suspension}
\begin{figure}
	\centering
	\includegraphics[width=0.75\columnwidth]{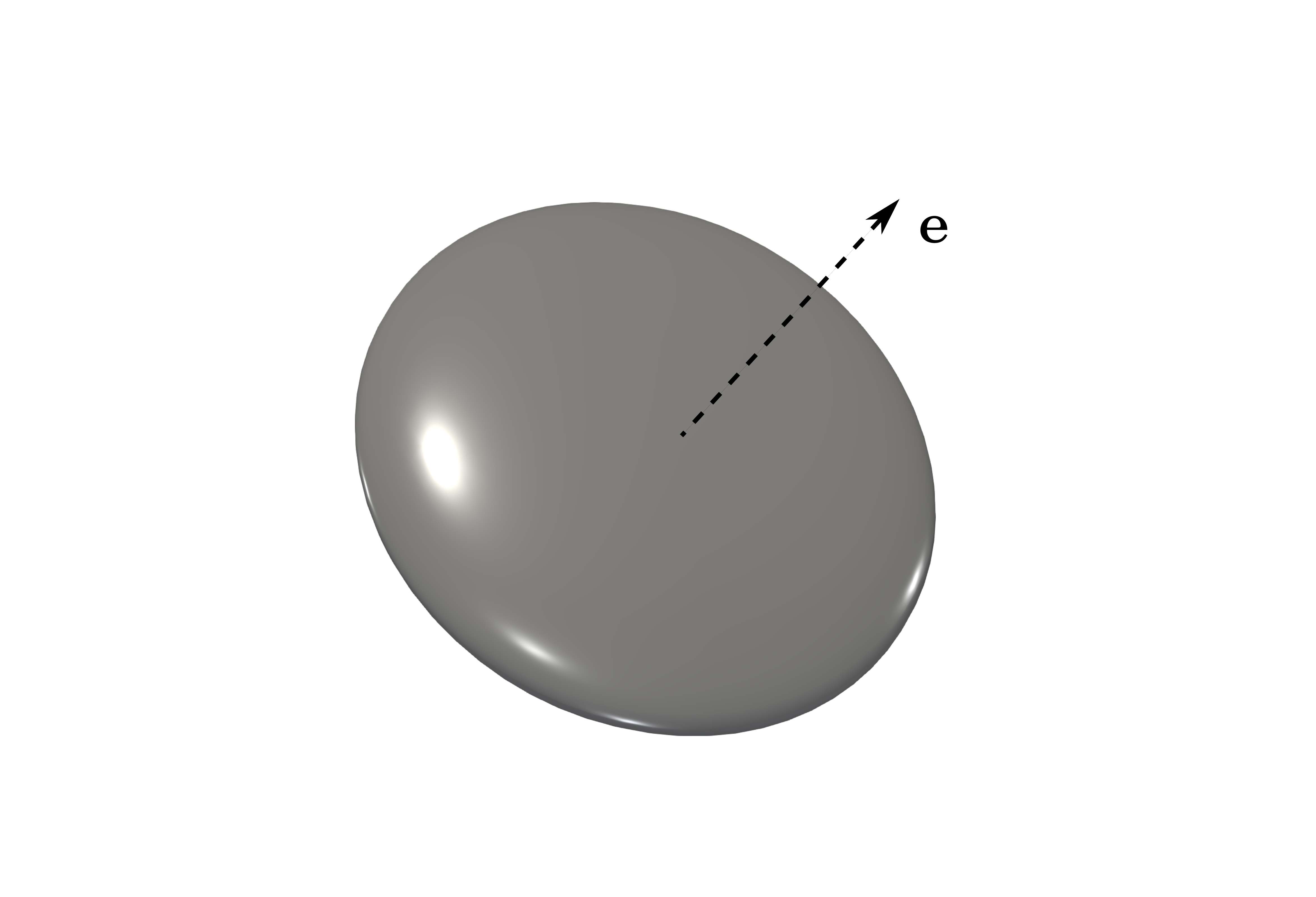}
	\caption{Oblate particle with symmetry axis \textbf{e}.}
	\label{fig:oblate_geometry}
\end{figure}
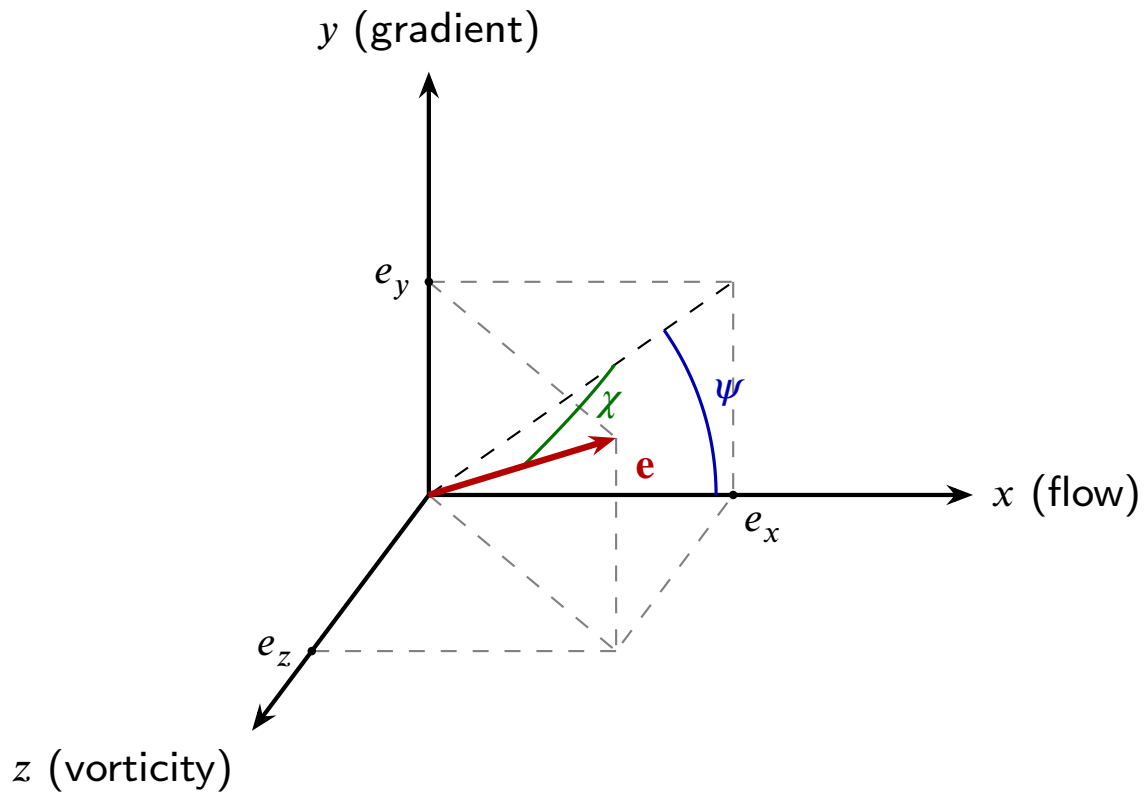
\begin{figure}
  \centering
  \begin{tikzpicture}[
    x={(1cm,0cm)}, y={(0cm,1cm)}, z={(-0.45cm,-0.6cm)},
    >={Stealth[length=5pt]}, font=\footnotesize]
    \pgfmathsetmacro{\PSI}{35}
    \pgfmathsetmacro{\CHI}{35}
    \pgfmathsetmacro{\R}{3}
    \pgfmathsetmacro{\ex}{\R*cos(\CHI)*cos(\PSI)}
    \pgfmathsetmacro{\ey}{\R*cos(\CHI)*sin(\PSI)}
    \pgfmathsetmacro{\ez}{\R*sin(\CHI)}
    \pgfmathsetmacro{\TH}{acos(cos(\CHI)*cos(\PSI))}
    
    \draw[->, thick] (0,0,0) -- (3.6,0,0) node[right] {$x$ (flow)};
    \draw[->, thick] (0,0,0) -- (0,2.8,0) node[above] {$y$ (gradient)};
    \draw[->, thick] (0,0,0) -- (0,0,2.6) node[below left] {$z$ (vorticity)};
    
    \draw[dashed, gray] (\ex,\ey,0) -- (\ex,0,0);
    \draw[dashed, gray] (\ex,\ey,\ez) -- (\ex,0,\ez) -- (\ex,0,0);
    \draw[dashed, gray] (\ex,0,\ez) -- (0,0,\ez);
    \draw[dashed, gray] (0,0,0) -- (\ex,0,\ez);
    \draw[dashed, gray] (0,\ey,0) -- (\ex,\ey,0);
    \draw[dashed, gray] (0,\ey,0) -- (\ex,\ey,\ez);
    \draw[dashed] (0,0,0) -- (\ex,\ey,0);
    
    \node[below] at (\ex+0.2,0,0) {$e_x$};
    \node[left]  at (0,\ey,0) {$e_y$};
    \node[left]  at (0,0,\ez) {$e_z$};
    \foreach \p in {(\ex,0,0),(0,\ey,0),(0,0,\ez)} \fill \p circle (0.8pt);
    
    \draw[blue!70!black, semithick] plot[domain=0:\PSI, samples=30]
      ({1.9*cos(\x)}, {1.9*sin(\x)}, 0);
    \node[blue!70!black, right] at ({1.9*cos(0.6*\PSI)}, {1.9*sin(0.6*\PSI)}, 0) {$\psi$};
    \draw[green!45!black, semithick] plot[domain=0:\CHI, samples=30]
      ({1.5*cos(\x)*cos(\PSI)}, {1.5*cos(\x)*sin(\PSI)}, {1.5*sin(\x)});
    \node[green!45!black] at
      ({1.85*cos(0.5*\CHI)*cos(\PSI)}, {1.7*cos(0.5*\CHI)*sin(\PSI)}, {1.85*sin(0.5*\CHI)}) {$\chi$};
    \draw[->, very thick, red!70!black] (0,0,0) -- (\ex,\ey,\ez)
      node[below right] {$\mathbf{e}$};
  \end{tikzpicture}
  \caption{Orientation of the symmetry axis $\mathbf{e}$ of an oblate particle. In blue, $\psi$ is the rotation of $\mathbf{e}$ within the shear plane $(x,y)$, in green $\chi$ its tilt out of the shear plane.}
 \label{fig:angles}
  \end{figure}
The focus of the current work was on ellipsoids resembling blood platelets, as stated in Section \ref{sec:intro}. They were modeled as oblate particles of aspect ratio $\lambda=0.25$ with polar and equatorial semi-axes \textit{c} and \textit{a} respectively, defining  $\deq=2(a^2c)^{1/3}$ the diameter of the volume equivalent sphere. On the lattice, the equatorial diameter spans $31$ cells, the polar axis $8$ cells, while the volume-equivalent sphere diameter $20$. Additional details regarding the chosen particle resolution and its validity for the current application will be given in Section \ref{sec:valid}.

Oblate particles are axisimmetric with respect to their polar semi-axis \textit{c} and their spatial orientation is described by the unit vector $\mathbf{e}$ along the symmetry axis (Figure \ref{fig:oblate_geometry}),
\begin{equation}
\mathbf{e}=\left(\cos\chi\cos\psi,\;\cos\chi\sin\psi,\;\sin\chi\right),
\label{eq:orientation}
\end{equation}
where $\psi$ is the rotation of the axis within the shear plane and $\chi$ its tilt out of it, as depicted in Figure \ref{fig:angles}. Since $\mathbf{e}$ and $-\mathbf{e}$ describe the same particle, the sign was chosen so that $e_x\ge0$, and the incidence angle $\theta$ between the flow and the symmetry axis follows from $\cos\theta=e_x$, with $\theta=0^\circ$ for an oblate face-on to the flow and $\theta=90^\circ$ in the edge-on case.

The suspensions were designed so that the oblate particles would be randomly oriented in space, therefore reflecting the target application: platelets in blood flow. Indeed, these blood cells do not align at a fixed angle, on the contrary they constantly collide with other blood cells as they travel towards the vessel walls. To mimic this disorder, the particle placement at each simulated solid volume fraction $\phi$, was generated by random sequential addition. Particles were inserted one at a time, each with a position drawn uniformly within the domain and an orientation drawn uniformly over all rotations. A candidate particle was accepted only if it did not overlap any previously placed particle, of course accounting for the periodic images in the streamwise and spanwise directions. The resulting suspensions were isotropic at every $\phi$, including the most dilute cases.

An additional check on the isotropy of each generated suspension was performed a posteriori by computation of the nematic order tensor \textit{S} \cite{advani1987}, defined as 
\begin{equation}
    S = \langle \mathbf{e}\mathbf{e}^{T} \rangle.
\end{equation}
Given a certain $\phi$, hence a determinate number of particles \textit{N}, a value of $\textit{S}$ close to its random sample value for the same \textit{N}, ensured that the simulated suspension is isotropic. Finally, the resulting positions and orientations were frozen before the shear flow was imposed, as is standard practice in particle-resolved studies of assemblies \cite{sanjeevi2020,li2021oblate,cao2020}, so that the loads can be referred to a prescribed position, orientation and slip.

\subsection{Forces and torque extraction}\label{sec:methods:loads}
The hydrodynamic loads exerted by the fluid on each particle were computed by means of the momentum exchange method \cite{ladd1994,mei2002}, which is consistent with the bounce-back treatment of the fluid-particle interface. Forces $\mathbf{F}$ were normalized by the Stokes drag of the volume-equivalent sphere and torques $\mathbf{T}$ by the scale of the Jeffery torque,
\begin{equation}
\mathbf{F}_{\mathrm{norm}}=\frac{\mathbf{F}}{3\pi\mu\,d_{\mathrm{eq}}\,U},\qquad
\mathbf{T}_{\mathrm{norm}}=\frac{\mathbf{T}}{\mu\,\dot\gamma\,d_{\mathrm{eq}}^3},
\label{eq:norm}
\end{equation}
with $\mu$ the dynamic viscosity, $d_{\mathrm{eq}}$ the diameter of the volume-equivalent sphere, $\dot\gamma$ the achieved shear rate and $U=\dot\gamma h$ the undisturbed velocity at the height $h$ of the platelet center, which defines the slip Reynolds number $\Rey=U\deq/\nu$. The normalized force is decomposed along the flow direction \textit{x} into the particle drag $D$ and lift components $L_y$ and $L_z$, along the gradient and the vorticity directions respectively:
\begin{equation}
D=\mathbf{F}_{\mathrm{norm}}\cdot\hat{\mathbf{x}},\qquad
L_y=\mathbf{F}_{\mathrm{norm}}\cdot\hat{\mathbf{y}},\qquad
L_z=\mathbf{F}_{\mathrm{norm}}\cdot\hat{\mathbf{z}}.
\label{eq:dl}
\end{equation}
In the Stokes limit, the force on a fixed spheroid depends on the flow only through the slip, and the components of $\mathbf{F}_{\mathrm{norm}}$ follow from its resistance tensor $f_\perp\mathbf{I}+\left(f_\parallel-f_\perp\right)\mathbf{ee}$
  \cite{happel1983,kim1991}:
\begin{align}
  D   &=f_\perp+\left(f_\parallel-f_\perp\right)e_x^2, \label{eq:dlz_stokes}\\
  L_y &=\left(f_\parallel-f_\perp\right)e_xe_y,\\
  L_z &=\left(f_\parallel-f_\perp\right)e_xe_z,
\end{align}
where $f_\parallel=1.3768$ and $f_\perp=1.0831$ are the drag factors for flow along and across the symmetry axis of an oblate particle of aspect ratio 0.25. In these conditions, the transverse force is purely orientation-induced and lies in the plane of $\hat{\mathbf{x}}$ and $\mathbf{e}$ with magnitude $\left(f_\parallel-f_\perp\right)\sin\theta\cos\theta$, which is the lift usually defined for spheroids in uniform flow \cite{zastawny2012,sanjeevi2018}. Projecting it on the gradient and vorticity directions instead isolates $L_y$, the only component in which a shear-induced lift can arise \cite{saffman1965,harper1968}. Similarly, the torque depends only on the shear, and the components of $\mathbf{T}_{\mathrm{norm}}$ are those of the Jeffery torque \cite{jeffery1922}:
\begin{align}
  T_x &=\tfrac12\left(Y-X-Z\right)e_xe_z, \nonumber \\
  T_y &=\tfrac12\left(Y-X+Z\right)e_ye_z, \label{eq:jeffery} \\
  T_z &=-\tfrac12Y+\tfrac12Z\left(e_x^2-e_y^2\right)-\tfrac12\left(X-Y\right)e_z^2, \nonumber
\end{align}
where $X=7.0671$ and $Y=5.7903$ are the rotational resistances of the above mentioned oblate about its symmetry and equatorial axis respectively, whereas $Z=Y(\lambda^2-1)/(\lambda^2+1)=-5.1091$ couples the torque to the strain. $T_z$ drives the rotation along the Jeffery orbit, while $T_x$ and $T_y$ appear only when the axis leaves the shear plane.

These expressions are exact for an isolated oblate ellipsoid at vanishing Reynolds number. The closure developed in this work is built around them, retaining their dependence on orientation $\mathbf{e}$ and using them as reference against which the effects of inertia and crowding will be measured, so that as $Re\to0$ and $\phi\to0$, it reduces to Eqs.~\eqref{eq:dlz_stokes}--\eqref{eq:jeffery}.

\subsection{Simulations setup}\label{sec:setup}
In the present study, each simulation was performed in the context of a cubic domain of side $L=31d_{eq}$ corresponding to 600 lattice nodes. The streamwise and spanwise boundaries were chosen periodic, while a free-slip boundary condition was prescribed on the walls normal to the velocity gradient, as schematically depicted in Figure \ref{fig:domain}. 

The fluid density and kinematic viscosity were those of blood plasma, namely $\rho=1025\,kg/m^3$ and $\nu=1.17\times10^{-6}\,m^2/s$. The relaxation time was kept at $\tau=1$, giving a time step $\Delta t=1.395\, ns$. The flow was initialized at rest and the target profile was gradually ramped from zero to its final value over the first $5\,000$ time steps so as to avoid the shock of an impulsively imposed shear.

The regime of interest covered the range $0.1\le\Rey\le1$, with $\Rey$ the slip Reynolds number defined in Section \ref{sec:methods:loads}. Since the slip $U=\dot\gamma h$ grows linearly in \textit{y}, a single shear rate would span only a limited range of $\Rey$ over the adopted domain instead of the full range under study. Therefore, as enlarging the domain at the same resolution was computationally out of reach, each volume fraction was simulated at three shear rates, set through the slip Reynolds number at the top of the domain as $\dot\gamma=Re_{\mathrm{top}}\,\nu/(\deq L)$ with $Re_{\mathrm{top}}=0.32$, $1.0$ and $1.47$, so that the oblates of the three groups together covered the whole regime of interest.

Not every platelet was retained in the analysis. Platelets whose center lied within $6\,\deq$ of either free-slip plane, in the gray bands of Figure~\ref{fig:domain}, were excluded to keep the closure free of the influence of the boundaries. The choice of this distance stems from the study of Zhang et al.\ \cite{zhang2026} who found wall effects on an oblate spheroid of the same aspect ratio negligible for a surface gap above $3.5\,\deq$. Its adequacy will be further verified in Section \ref{sec:res_suspensions}. In addition, particles whose surface lies closer than $1.5$ lattice cells to that of a neighbor were also excluded, since the flow within such narrow gaps is not resolved. The analyzed population comprises 15 seeded configurations per each of the denser volume fractions $\phi=\left\{0.01,0.02,0.05,0.075,0.10,0.15\right\}$ and 27 seeded configurations per each of the three dilute volume fractions $\phi=\left\{0.002,0.004,0.008\right\}$ for a total of 171 simulations and 171'079 retained particles. Each suspension was initialized via a unique seed of a random generator, so that no two simulations would share the same particle arrangement. An example is given in Figure \ref{fig:setup_3d}.
\begin{figure}
\centering
\begin{tikzpicture}[x=0.5cm, y=0.5cm, >=Stealth, font=\footnotesize]

  \fill[gray!20] (0,0)   rectangle (10,2.6);
  \fill[gray!20] (0,7.4) rectangle (10,10);

  \draw (0,0) rectangle (10,10);
  \draw[dashed] (0,2.6) -- (10,2.6);
  \draw[dashed] (0,7.4) -- (10,7.4);

  \draw[<->] (8.3,0)   -- node[right] {$6d_{\mathrm{eq}}$} (8.3,2.6);
  \draw[<->] (8.3,7.4) -- node[right] {$6d_{\mathrm{eq}}$} (8.3,10);

  \node[below] at (5,10) {Free-slip};
  \node[above] at (5,0)  {Free-slip};
  \node[rotate=90, above] at (0,5)  {Periodic};   %
  \node[rotate=90, below] at (10,5) {Periodic};   %

  \draw[->] (3,10.5) -- node[above] {$U=\dot{\gamma}\,y$} (7,10.5);

  \node[below] at (5,-0.3) {$U=0$};

  \path (10.7,5) (5,-2.3);

  \begin{scope}[overlay, xshift=1.2cm]
    \draw[->] (-4.2,-1.6) -- (-4.2,0.6) node[above] {$Y$};
    \draw[->] (-4.2,-1.6) -- (-2.2,-1.6) node[right] {$X$};
    \draw (-4.2,-1.1) -- (-3.7,-1.1) -- (-3.7,-1.6);
    \node[below left] at (-4.2,-1.6) {$Z$};
  \end{scope}

\end{tikzpicture}
\caption{Computational domain and boundary conditions.}
\label{fig:domain}
\end{figure}
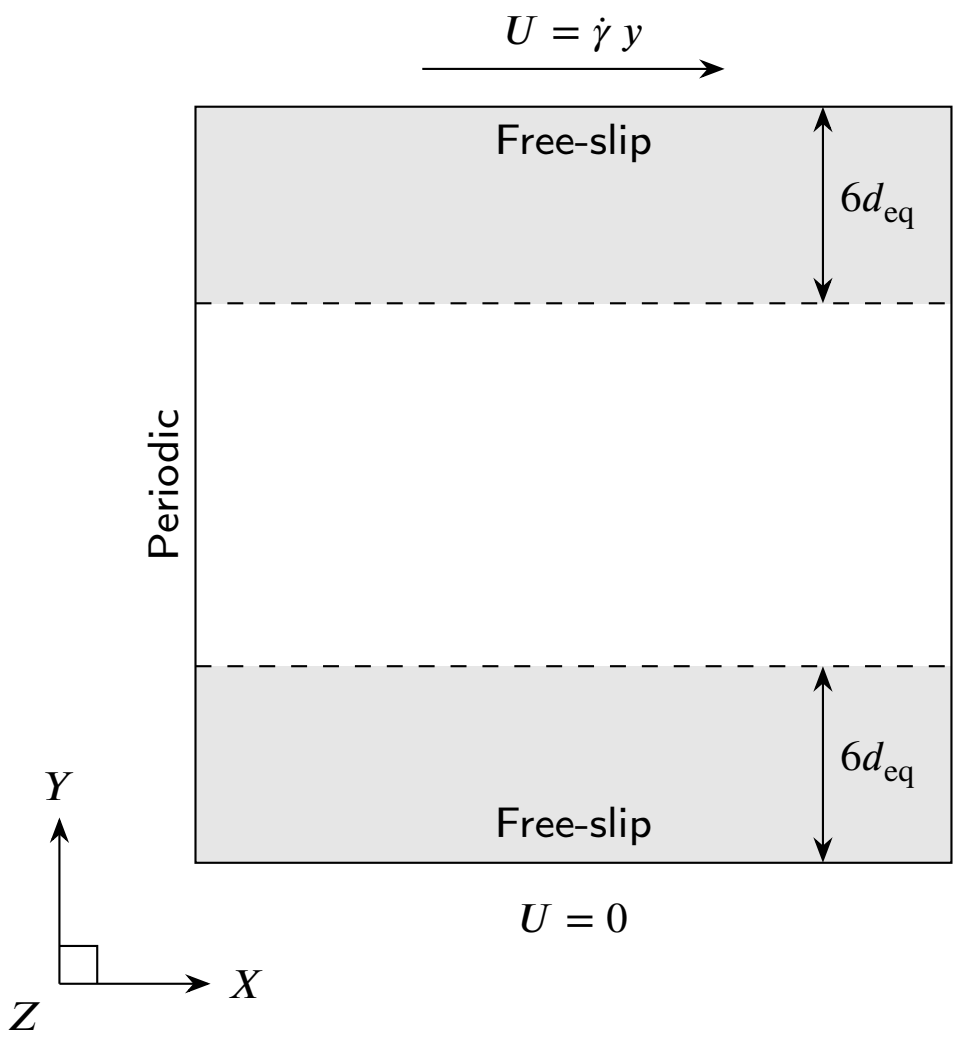
\begin{figure}
	\centering
	\includegraphics[width=1\columnwidth]{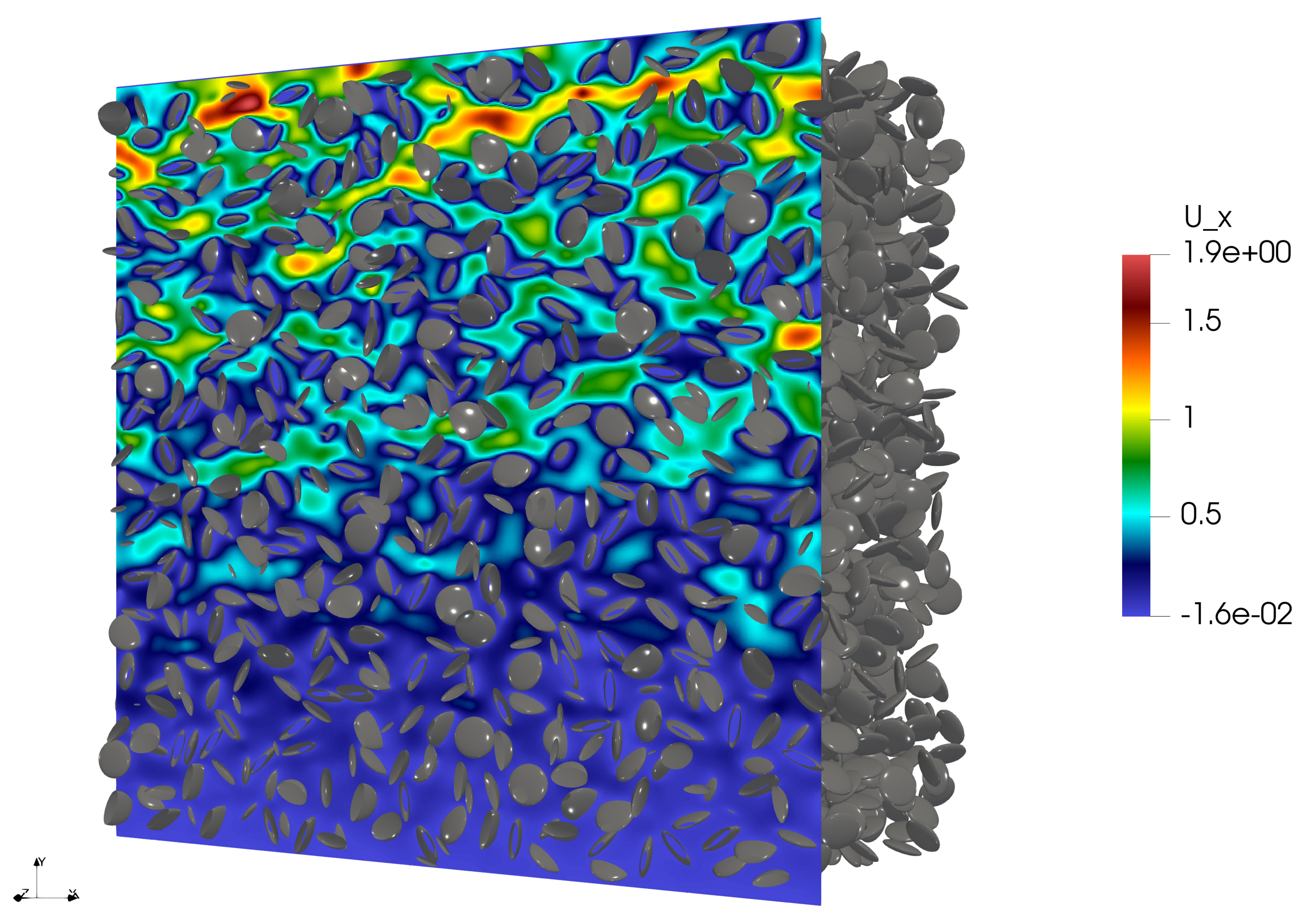}
	\caption{Example cross section of the simulation set up, here at $\phi=15\%$. In color: velocity field for the $x$ component of \textbf{U}.}
	\label{fig:setup_3d}
\end{figure}

\section{Results and discussion}\label{sec:results}
The following paragraphs will focus on reporting how drag, lift and torque felt by an oblate ellipsoid in suspension are impacted by oblate orientation, neighboring particles and flow conditions via the exploration of a range of Reynolds numbers and solid volume fractions. After a brief validation of the solver, single oblate simulations lay the foundation for the closure providing the isolated limit values, at last simulations of suspensions are discussed.
\begin{figure}
  \centering
  \includegraphics[width=0.95\columnwidth]{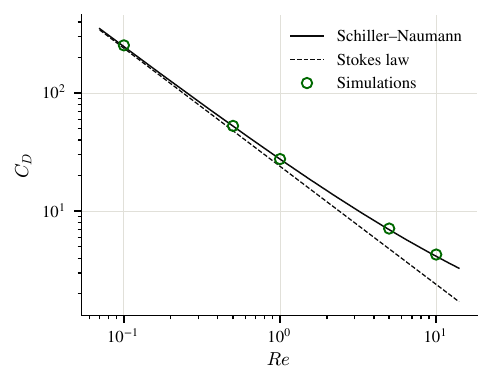}
  \caption{Drag coefficient of a sphere in uniform flow. Simulations (green circles) against the correlation of Schiller-Naumann \cite{schiller1933} and the Stokes drag.}
  \label{fig:valid_sphere}
\end{figure}

\subsection{Validations}\label{sec:valid}
As stated in Section \ref{sec:method}, the code in use is an extension of the LEDDS solver \cite{rambosson2026},introducing the usage of the Bouzidi condition at the fluid-particle interface and a forcing term to drive the fluid flow. Consequently, it was imperative for us to validate this updated version against well known results to prove the reliability of the solver and at the same time demonstrate the accuracy of the employed grid resolution. 
\paragraph{Sphere in uniform flow.} The drag coefficient $C_D$ values for flow around a sphere at increasing $\Rey$ are shown in Figure \ref{fig:valid_sphere} and compared with Schiller-Naumann's correlation \cite{schiller1933}. The sphere of diameter $d=d_{eq}$ spanning 20 cells, was placed fixed at the center of a cubic computational domain identical in size and resolution to the one later adopted for the suspension simulations described in Section \ref{sec:setup}, and exposed to a uniform flow of velocity $U$ such that $Re=\rho U d/\mu=0.1$, $0.5$, $1$, $5$ and $10$. The results from our simulation are in good agreement with not only the community standard of Schiller-Naumann, but also the existing literature \cite{sanjeevi2018}. This outcome ensures the ability of the solver to reproduce known results and proves the chosen resolution sufficient to fully resolve the fluid dynamics around a particle in Stokes regime. Indeed, employing 20 cells per diameter places us safely above the threshold indicated by Sanjeevi et al. \cite{sanjeevi2018} in their convergence and grid resolution study, an investigation carried out with the same numerical method and boundary conditions as the present work.
\begin{figure}
  \centering
  \includegraphics[width=0.9\columnwidth]{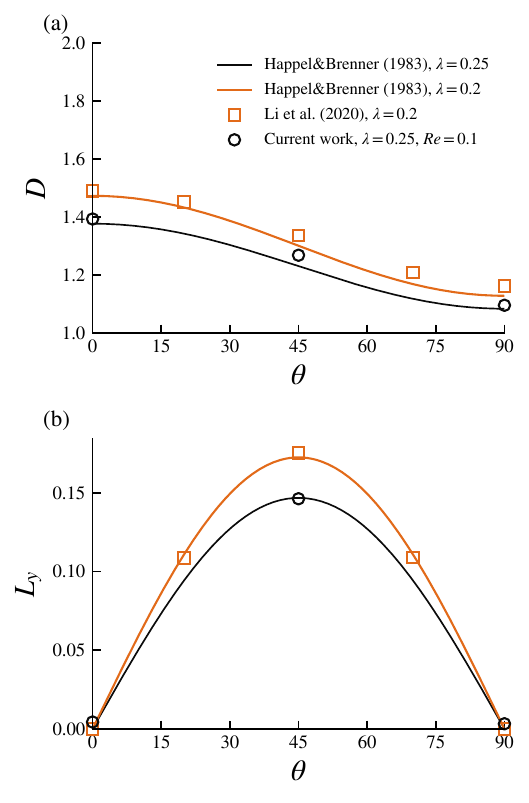}
  \caption{Normalized drag (a) and lift (b) on an isolated oblate at mid-height as function of the incidence angle $\theta$. Simulations at $\Rey=0.1$ (black circles) are compared with the work of Li et al.\ \cite{li2021oblate} for $\lambda=0.2$ at $\Rey<0.1$  (Orange squares). Lines are unbounded Stokes drag and lift for an oblate in uniform flow \cite{happel1983}.}
  \label{fig:valid_single_oblate}
\end{figure}

\paragraph{Oblate in shear flow.} To further corroborate the adequacy of the solver in modeling suspensions of oblate particles, a single oblate ellipsoid of aspect ratio $\lambda=0.25$ as described in Section \ref{sec:suspension} was fixed at the center of the computational cubic domain in use and invested by a linear shear flow. Three different orientations were explored, prescribing $\theta=0^\circ$, $45^\circ$ and $90^\circ$, the angle between the slip and the particle symmetry axis. Figure \ref{fig:valid_single_oblate} displays the resulting drag and lift forces acting on the particle as a function of $\theta$, together with Happel\&Brenner's correlation \cite{happel1983}. It must be noted that the forces presented here have been corrected to account for the particle's own periodic image. Indeed, when comparing numerical values with the unbounded Stokes solution, the periodicity of the computational domain needs to be factored in. A particle in a periodic domain as the one employed here, feels its own images thereby experiencing an increase in its own drag, as described by Hasimoto \cite{hasimoto1959}. This effect can be corrected by multiplying the simulated forces by the factor $1-2.837\ell/L$, where $\ell=f_k\deq/2$ is the Stokes radius of the particle along the direction of motion and $f_k$ is the corresponding Stokes drag factor.

The corrected values show very good agreement with the unbounded Stokes reference, confirming once more the adequacy of the resolution as well as the solver's ability to accurately capture the hydrodynamic response of oblate particles. Interestingly, although the correlation by Happel\&Brenner was originally derived for spheroids in uniform flow, in the Stokes regime, it remains applicable to shear flow configurations as the one at hand. Indeed, in the case of a fixed centrosymmetric particle like the oblate at study, the hydrodynamic force solely depends on the local slip velocity, as the straining component of the shear flow offers no contribution \cite{faxen1922,brenner1964}. The drag is therefore equivalent to the one experienced in a uniform stream with velocity $U=\dot\gamma h$, with $h$ the particle's distance from the bottom wall. The results are also consistent with the simulations of Li et al. \cite{li2021oblate}, with the small discrepancies to be attributed solely to the different aspect ratio considered in their study ($\lambda=0.2$).

\subsection{Isolated oblate}\label{sec:isolated_oblate}
To verify that the closure would indeed tend to the exact Stokes loads of an isolated particle and allow the effects of slip and shear to be separated, simulations of a single oblate were set up under two configurations. First, the particle was kept fixed in the middle of the domain and placed face-on (axis along $x$) and edge-on to the flow (axis along $y$ and $z$) and tilted by $\psi=45^\circ$ within and $\chi=45^\circ$ out of the shear plane with $\Rey=0.1$, $0.5$ and $1$ imposed at height of the particle center. Afterwards, the particle was kept tilted either by $\chi=45^\circ$ or $\psi=45^\circ$ and placed at $h=4.8$, $8.1$, $13.9$, $15.5$ and $23.8\,\deq$ from the bottom wall, at the same three $\Rey$ at particle center. With this second series of simulations it was possible to change the shear rate relative to the slip while keeping the slip $\Rey$ constant, indeed $\deq\dot\gamma/U=\deq/h$, therefore analyzing the effects of slip and shear separately.
\paragraph{Forces.} At mid-height, the drag of the oblate ellipsoid tilted by $45^\circ$ within or out of the shear plane was found to come within $0.5\%$ of the drag computed as the average between face-on and edge-on values at every considered $\Rey$, thus confirming the requirement of Eq.~\eqref{eq:dlz_stokes}. In addition, the drag force grew between $\Rey=0.1$ and $1$ by $9\%$ and $7\%$ respectively for face-on and edge-on orientations. The measured increase is to be attributed to inertia, but because of the periodic domain it was far smaller than the one expected from the prediction made through the unbounded Oseen correction \cite{brenner1961}. To be precise, in Stokes flow the disturbance created by a particle decays slowly with distance, but if the domain is periodic it reaches its periodic images, raising the drag. As $\Rey$ increases, the periodic images lose part of their effect once they lie beyond Oseen length $\deq/\Rey$, marking the drag increase purely as an inertial effect.

The Stokes relation between lift and drag anisotropy was also found to hold over the whole range of $\Rey$. Indeed, the amplitude of the transverse force, namely $L_y/(e_xe_y)$ at $\psi=45^\circ$ and $L_z/(e_xe_z)$ at $\chi=45^\circ$, matched within $0.8\%$ the difference between face-on and edge-on drag, allowing the prediction of lift directly from drag alone without the need of a separate lift constant. To be exact, for the oblate tilted within the shear plane, the departure was found to increase from $0.6\%$ at $\Rey=0.1$ to $7.9\%$ at $\Rey=1$, an effect attributed to inertia. The regime of interest for the present work is precisely the one for which this drag-lift relation can be relied upon, as expressed by Sanjeevi et al.\ \cite{sanjeevi2018}.

Lastly, raising the oblate particle inside the simulation domain allowed to cover a gap left open by near-wall studies on oblate spheroids introduced in Section \ref{sec:intro}: whether shear rate acts on the force beyond the local slip. At fixed $\Rey$, moving the oblate between $h=4.8$ and $23.8\,\deq$ changed $\deq\dot\gamma/U$ fivefold while the drag by at most $0.9\%$. The force on the particle therefore depends on the slip alone and no explicit shear-rate term is needed in the force closure for $\Rey\le1$, in line with the Stokes result of Section \ref{sec:intro} and with the vanishing shear correction to the drag of Kurose\&Komori for spheres at $\Rey=1$ \cite{kurose1999} and of Chéron et al.\ for rods below $\Rey=1$ \cite{cheron2024a}. 

\paragraph{Torque.}
\begin{figure}
  \centering
  \includegraphics[width=0.99\columnwidth]{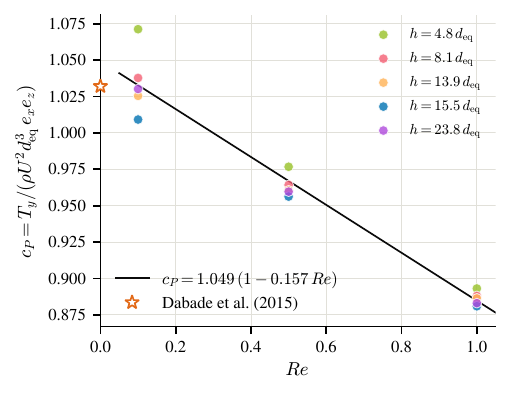}
  \caption{Pitching coefficient of an isolated oblate tilted by $\chi=45^\circ$, at five heights and three Reynolds numbers. Line is Eq. \eqref{eq:cp}. Star is theoretical Stokes limit \cite{dabade2015}.}
  \label{fig:pitching}
\end{figure}
For an oblate ellipsoid tilted by $\chi=45^\circ$ out of the shear plane, the Jeffery torque around $y$ should vanish. On the contrary, the computed $T_y$ was found to grow with the square of the slip. As this contribution was observed solely for the gradient direction of torque, the following form was adopted for its description:
\begin{equation}
\mathbf{T}_P=c_P\,\rho\,U^2\deq^3\,e_x\left(\mathbf{e}\times\hat{\mathbf{x}}\right),
\label{eq:pitching}
\end{equation}
The measurement of $c_P$ over the five heights tested in the second configuration highlighted a collapse of the coefficient to within $3.5\%$ at $\Rey=0.1$, $1.4\%$ at $0.5$ and $0.8\%$ at $1$ against a fivefold change in shear rate, as shown in Figure \ref{fig:pitching}. It is therefore evident how $c_P$ is governed by the slip and not by the shear, making $\mathbf{T}_P$ the pitching torque that turns a translating spheroid broadside-on \cite{cox1965,khayat1989} and
\begin{equation}
c_P=1.049\left(1-0.157\,\Rey\right),
\label{eq:cp}
\end{equation}
the coefficient of inertial pitching, linearly fitted over the results of Figure \ref{fig:pitching}. For $\Rey\to0$ the correlation is just $1.7\%$ above the theory of Dabade et al.\ for an aspect ratio of $0.25$ \cite{dabade2015}. Finally, once normalized as in Eq.~\eqref{eq:norm}, the pitching torque reads $c_P\,\Rey\,(h/\deq)\,e_x\left(\mathbf{e}\times\hat{\mathbf{x}}\right)$, so that at $\Rey=1$ and mid-height it exceeds the Jeffery torque of the tilted oblate.

\subsection{Suspensions}\label{sec:res_suspensions}
\begin{figure}
  \centering
  \includegraphics[width=0.99\columnwidth]{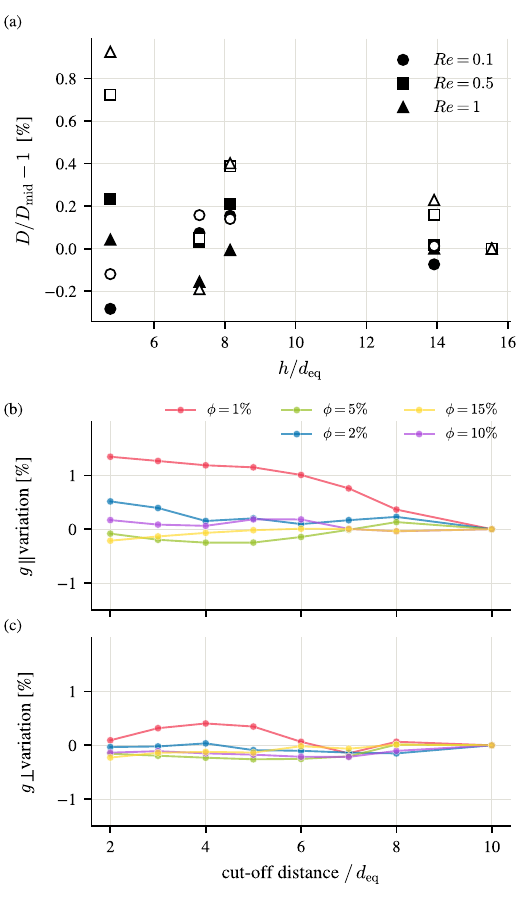}
  \caption{Influence of the free-slip planes. (a) Drag of the isolated oblate tilted by $\chi=45^\circ$ at $\Rey=0.1$, $0.5$ and $1$, relative to its value at mid-height, against the distance from the nearest plane. (b, c) Variation with the cut-off distance of suspension drag factors along ($g_\parallel$, b) and across ($g_\perp$, c) the symmetry axis, relative to their value at $10\,\deq$, for $\phi=1$, $2$, $5$, $10$ and $15\%$.}
  \label{fig:wallcut}
\end{figure}
\begin{figure*}
  \centering
  \includegraphics[width=0.99\textwidth]{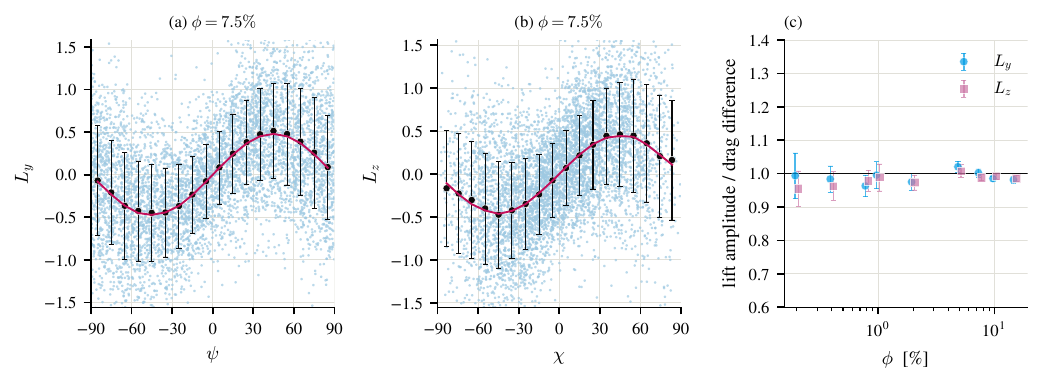}
  \caption{Lift at $\phi=7.5\%$ in the gradient (a) and vorticity (b) directions. Blue dots are individual particles (at most $6\,000$ per panel). Black points are means over $10^\circ$ ranges, with standard deviation of the individual particles in each range. Red curves are the prediction of Eqs.~\eqref{eq:lift_closure_y}--\eqref{eq:lift_closure_z}. (c) Ratio of the lift amplitude to the difference between face-on and edge-on drag with standard error over the realizations.}
  \label{fig:lift}
\end{figure*}

A total 171 different simulations were run to cover the nine volume fractions of interest, ensuring at each one, a good sampling of the orientation space and Reynolds range. Once converged, a first screening of the results was performed to identify the distance from the free-slip walls beyond which the suspension dynamics could be considered boundary independent. Inheriting from the isolated oblate simulations of Section \ref{sec:isolated_oblate}, Figure \ref{fig:wallcut}(a) reports for oblate ellipsoids at increasing distance from either top or bottom plane, the amount of drag force exceeding the value measured on a particle at mid-height at $Re=0.1, 0.5$ and $1$. The excess decreased from $0.3$--$0.9\%$ at $4.8\,\deq$ to $0.05$--$0.2\%$ at $7.3\,\deq$, reaching $0.2\%$ from $13.9\,\deq$. A cut-off distance of $6\,\deq$ was adopted as the initial estimate, as it was considered safe without being overly conservative. Once the closure was drafted, an additional check was performed on the suspension data by recalculating the parameters $g_\parallel$ and $g_\perp$, explained in detail in the following paragraphs, at increasing cut-off distances, from $2$ to $10\,\deq$. As can be observed in Figures \ref{fig:wallcut}(b) and (c), for $\phi\ge2\%$ both parameters resulted insensitive to the boundary plane with a variation of at most $1\%$ throughout the range. Thus, data from the suspension confirmed the safety of the initial estimate, allowing the preservation of a statistically significant number of particles per volume fraction.
At each $\phi$, the particles retained for analysis were grouped according to their slip Reynolds number, in five logarithmically spaced ranges between $0.1$ and $1$. Unlike in simulations of uniform flow past fixed assemblies \cite{sanjeevi2020,cao2022}, here the slip Reynolds number is not a parameter of a run but instead varies with the height of the particle within every run because the velocity profile is linear in \textit{h}. The obtained $\Rey$ ranges were used to fit the drag factors and test for any dependence on $\Rey$, however said dependence was found to be at most of a few percentage points and disappearing in the dense suspensions, as will be presented in Figure \ref{fig:drag_factors_Re}. It follows that shown graphs and closure group all $\Rey$ ranges, unless otherwise specified.
\begin{figure*}
  \centering
  \includegraphics[width=0.99\textwidth]{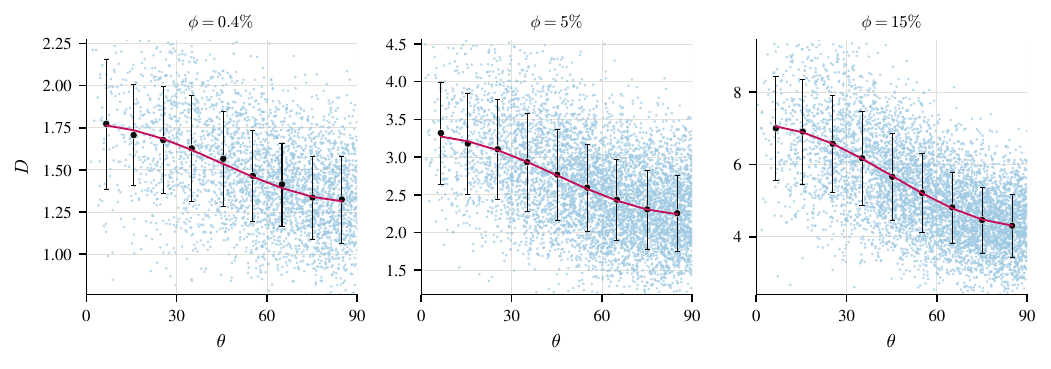}
  \caption{Normalized drag against incidence angle at $\phi=0.4$, $5$ and $15\%$, all $\Rey$ ranges together. Blue dots are individual particles (at most $6\,000$ per panel). Black points are means over $10^\circ$ ranges, with standard deviation of the individual particles in each range. Red curves are Eqs. \eqref{eq:drag_closure}--\eqref{eq:gk} evaluated on the same particles.}
  \label{fig:drag_theta}
\end{figure*}
Independently of their $\Rey$, but according to their orientation, particles were grouped in ranges of $10^\circ$ increments of each governing angle, namely $\theta$ for $D$, $\psi$ for $L_y$ and $T_z$, $\chi$ for $L_z$, $T_x$ and $T_y$.the extracted loads within each of the orientation groups were averaged and employed as direct comparison to the final closure. Figures \ref{fig:drag_theta} to \ref{fig:torque} display exactly these arithmetic means, along with the closure coefficients extrapolated by ordinary least squares on the individual particles. In addition, the inter particle variability induced by the neighbors was quantified by the standard deviation of the individual oblate loads within a group. This does not depend on the number of particle simulated, as it is a property of the suspension. Instead, the error bars of all the figures function of $\phi$ mark the standard deviation of a coefficient over the many independent realizations of a volume fraction, defining the quality of the coefficient estimation.

\subsubsection{Drag}\label{sec:res_drag}
\begin{figure}
  \centering
  \includegraphics[width=\columnwidth]{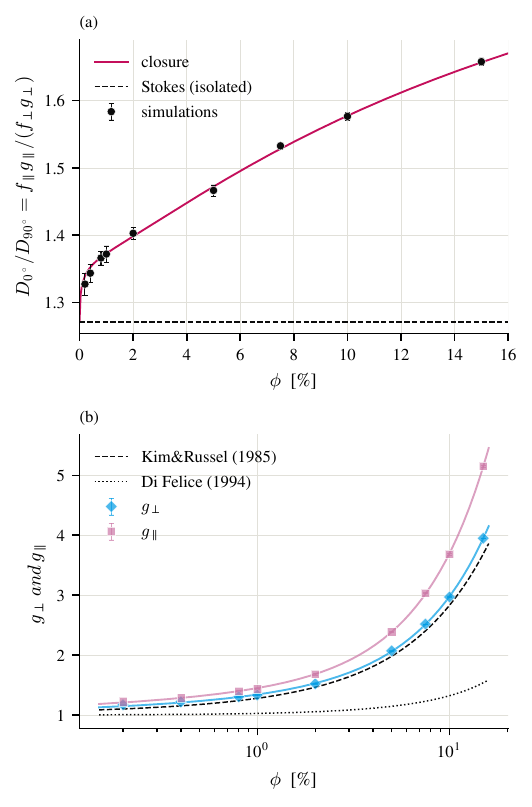}
  \caption{(a) Ratio of face-on to edge-on drag against volume fraction. Curve: closure. Dashed line: isolated oblate in Stokes flow. (b) Drag factors along ($g_\parallel$) and across ($g_\perp$) the symmetry axis against volume fraction. Points: simulations, with the standard error over the realizations. Solid curves: Eq.~\eqref{eq:gk}. Dashed black curve: Kim and Russel for spheres \cite{kimrussel1985}. Dotted curve: voidage function \cite{difelice1994}.}
  \label{fig:drag_factors}
\end{figure}
The drag force \textit{D} of individual particles extracted from multiparticle simulations is presented in Figure~\ref{fig:drag_theta} for three exemplificative volume fractions as a function of the incidence angle $\theta$. The use of a scatterplot facilitates in identifying two key features: the large inter-particle variability and the decrease in mean drag from face-on to edge-on orientation. Regarding the first, such a behavior was to be expected as a direct consequence of the wide diversity of neighborhoods of each particle in suspension and is well in accordance with what reported for assemblies of spherocylinders and prolate ellipsoids \cite{sanjeevi2020,cao2022}. The standard deviation within a $10^\circ$ range was measured at $15$--$22\%$ of the mean at every volume fraction and is independent of orientation and $\Rey$, indeed the five $\Rey$ ranges agree to within $3.5$ percentage points, without a trend. In terms of the second feature, it appeared clear how drag at every volume fraction scales as $\cos^2\theta$, following exactly the dependence of an isolated oblate, thus going beyond the results by Sanjeevi\&Padding for spherocylinders in uniform flow \cite{sanjeevi2020} and proving the persistence of this scaling in a crowd in linear shear flow.

The average drag experienced by an oblate ellipsoid in suspension under shear flow can therefore be expressed by Eq.(\ref{eq:dlz_stokes}) with the addition of two factors and reads:
\begin{equation}
D=f_\perp g_\perp+\left(f_\parallel g_\parallel-f_\perp g_\perp\right)e_x^2 ,
\label{eq:drag_closure}
\end{equation}
with $g_\perp$ the intercept and $g_\parallel$ the slope of a straight line fitted to the drag of individual particles against $e_x^2$ at each volume fraction. To quantify the influence of crowding, both factors were plotted as function of the solid volume fraction in Figure \ref{fig:drag_factors}. In panel (a) the ratio between face-on and edge-on drag reveals how crowding effects are not felt equally on the two orientations. If that were the case, the curve would have remained almost flat for any volume fraction, instead $g_\parallel$ always exceeds $g_\perp$ against a constant $1.27$ for the isolated oblate in Stokes flow. The same trend can be found within the simulations of aligned oblate arrays by Li et al.\ who report a ratio of $1.74$ at $\lambda=0.25$ and $\phi=10\%$ for a fully aligned array \cite{li2021oblate}, instead it is absent in the study on spherocylinders by Sanjeevi\&Padding, whose ratio of perpendicular to parallel drag stays constant with $\phi$ at $\Rey=0.1$ \cite{sanjeevi2020}, clearly marking this as a geometry specific effect. Moreover, Figure \ref{fig:drag_factors}(b) shows how $g_\perp$ and $g_\parallel$ grow monotonically with $\phi$ right from the most dilute suspensions with $g_\perp=1.17$ and $g_\parallel=1.22$ at $0.2\%$ increasing to $3.95$ and $5.15$ at $15\%$ respectively. This was to be expected for fixed beds like the ones under study, since neighboring particles act as a screen to fluid flow increasing drag with $\phi^{1/2}$ \cite{brinkman1947,kimrussel1985}. Interestingly, comparing $g_\perp$ and $g_\parallel$ with two spherical voidage functions as depicted in Figure \ref{fig:drag_factors}(b), showed that the drag force of oblates in suspension cannot be built directly from the isolated oblate drag multiplied by a voidage function developed for spherical particles, as it is instead common in Eulerian--Lagrangian simulations. This is because, a function such as the one by Di Felice \cite{difelice1994} used in the inertial regime, lacks the ability to quantify the screening effect of neighboring particles resulting in the under prediction of drag at every volume fraction and orientation. The $17$--$22\%$ increment at $\phi=0.2\%$ quantified in the present study extends to the dilute range the poor adequacy of Di Felice's law already reported by Li et al.\ at $\phi\ge10\%$ \cite{li2021oblate}. Alternatively, the comparison with Kim\&Russel's extension for random arrays of spheres at low $\Rey$, which contains the $\phi^{1/2}$ screening factor \cite{kimrussel1985}, showed $g_\perp$ following within $4$--$7\%$ at every volume fraction. This was not the case though for $g_\parallel$, which rises above Kim\&Russel's law by $10\%$ already at $\phi=0.2\%$ and keeps distancing itself as volume fraction increases. Moreover, past the $\phi\approx5\%$ limit, the expansion would be used beyond the range it was intended for. 
\begin{figure}
  \centering
  \includegraphics[width=\columnwidth]{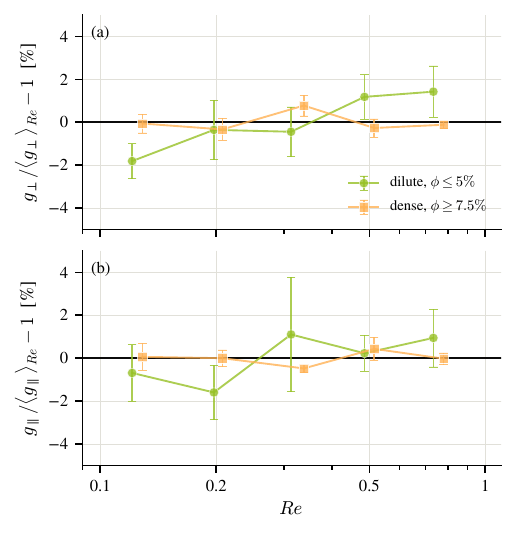}
  \caption{Drag factors: (a) across ($g_\perp$) and (b) along ($g_\parallel$) the symmetry axis in the five ranges of the slip Reynolds number, for the dilute and the dense suspensions. At each volume fraction, the factor of a range is obtained from the line of Eq.~\eqref{eq:drag_closure} fitted on its particles and is referred to the mean over the five ranges. Squares: average of these relative factors over the volume fractions of the group, at the mean $\Rey$ of the range (the two groups slightly shifted in $\Rey$ for legibility). Bars: their standard deviation over the volume fractions.}
  \label{fig:drag_factors_Re}
\end{figure}

This partial good accordance guided the decision to model the two factors in the likeness of the dilute expansion  \cite{kimrussel1985} such that
\begin{equation}
g_k=1+a_k\,\phi^{1/2}+b_k\,\phi\ln\phi+c_k\,\phi,\qquad k=\perp,\parallel,
\label{eq:gk}
\end{equation}
with $a_k$, $b_k$ and $c_k$ free coefficients the values of which are reported in Table \ref{tab:coeff}. The fit reproduces the factors of every volume fraction and $\Rey$ range with a root-mean-square deviation of $1.5\%$ for $g_\perp$ and $1.6\%$ for $g_\parallel$, and below $0.2\%$ it tends to the isolated oblate by construction.

Equations \ref{eq:drag_closure} and \ref{eq:gk} were designed without an explicit dependence on the Reynolds number, a characteristic that requires to be addressed. To investigate if the simulation data pointed towards such a dependence, the two factors $g_\perp$ and $g_\parallel$, computed for every volume fraction at every $\Rey$ range and referred to their mean over the five logarithmically spaced $\Rey$ ranges, were averaged separately for the dilute suspensions ($\phi\le5\%$) and the dense ones ($\phi\ge7.5\%$) and then plotted as in Figure \ref{fig:drag_factors_Re}, together with with their standard deviation. It can be observed how in the dilute suspensions, $g_\perp$ grows with $\Rey$ by about $3\%$ across the range, an increase larger than the spread between volume fractions which amount to $0.8$--$1.4\%$. On the contrary, the averaged $g_\parallel$ shows no trend, even changing sign between neighboring $\Rey$ ranges. As the volume fraction increases, the trend on $g_\perp$ disappears and both $g_k$ factors of dense suspensions are predominantly flat, pointing towards no clear $\Rey$ dependence. This behavior can be explained in terms of the ratio between the Brinkman length and the Oseen length. The former characterizes the distance over which momentum is screened by the surrounding particles, whereas the latter, $\deq/\Rey$, represents the length scale over which inertial effects become relevant. At low $\Rey$, the Oseen length is much larger than the Brinkman length, so that momentum is screened by the neighbors before inertia can affect the flow. As $\Rey$ increases, the Oseen length decreases and can become comparable to or smaller than the Brinkman length, allowing inertial effects to contribute to the drag. In the present study, the Brinkman length is $4.9\,\deq$ at $\phi=0.2\%$, scaling as $\phi^{-1/2}$ and reaching $2.2\,\deq$ at $\phi=1\%$ while falling below $\deq$ for $\phi\geq 5\%$. By contrast, the Oseen length ranges from $\deq$ to $10\,\deq$ over the investigated conditions. Thus, the Brinkman length exceeds the Oseen length over part of the range of interest only for the most dilute systems, consistent with an inertial contribution to the drag persisting in these cases. The dependence on $\Rey$ is therefore confined to the most dilute suspensions and adding a $\Rey$ term would lower the root-mean-square deviation of the fit only from $1.5$ to $1.2\%$ for $g_\perp$ and from $1.6$ to $1.5\%$ for $g_\parallel$, at the cost of an additional constant, and was therefore not included.

\subsubsection{Lift}\label{sec:res_lift}
\begin{figure}
  \centering
  \includegraphics[width=\columnwidth]{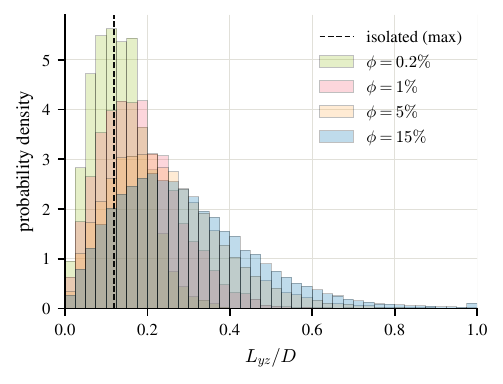}
  \caption{Distribution of the magnitude of the lift relative to the drag on the individual particles, $L_{yz}/D$ with $L_{yz}=(L_y^2+L_z^2)^{1/2}$, at four volume fractions, all $\Rey$ ranges together. Dashed line: maximum for the isolated oblate in Stokes flow, at $45^\circ$.}
  \label{fig:lift_magnitude}
\end{figure}
In Section \ref{sec:isolated_oblate} the lift of an isolated oblate particle in shear flow was directly derived from drag without any ad hoc lift parameter. To evaluate whether the same can be said in the case of oblates in suspension, the transverse forces in the gradient and vorticity direction of individual particles extracted from multiple simulations were plotted for $\phi=7.5\%$ in Figure \ref{fig:lift}, superposing on the scatter plots the hypothesized closure equation of the form  
\begin{align}
L_y&=\left(f_\parallel g_\parallel-f_\perp g_\perp\right)e_xe_y, \label{eq:lift_closure_y} \\
L_z&=\left(f_\parallel g_\parallel-f_\perp g_\perp\right)e_xe_z, \label{eq:lift_closure_z}
\end{align}
where the prefactor is taken from the drag, without fitting anything to the cloud of points. Panel (a) already shows a close agreement between the simulation values and the closure, a result corroborated by panel (c) in which the ratio between the lift amplitude and the slope of the drag in $e_x^2$ as a function of $\phi$ equals to around 1 for any volume fraction, with accuracy getting better as $\phi$ increases. This result captures, within inter particle scatter, how lift on an oblate in the suspension is the projection of its drag anisotropy in both transverse directions. Specifically, the ratio gave $0.97$--$1.02$ for $\phi\ge2\%$ and $0.96$--$1.00$ below, where fewer particles are available. As for the drag, Figure \ref{fig:lift} (a) and (b) reveals that the individual particles scatter around the mean, with a standard deviation comparable to the lift amplitude. Cao\&Tafti found the same relation to hold for suspensions of prolate ellipsoids of aspect ratio $2.5$ up to $\Rey=200$, but for aspect ratio $10$ only at $\Rey=10$, concluding that it applies to particles of moderate anisotropy at low $\Rey$ \cite{cao2022}, indeed the present oblate ellipsoid fall within this regime.

When it comes to the application of hydrodynamic closures in CFD-DEM simulations, lift can be sometimes neglected depending on its magnitude relative to the drag on each particle \cite{sanjeevi2020}. Figure \ref{fig:lift_magnitude} explores for four volume fractions exemplificative volume fractions how probable it is for the magnitude of the transverse force, namely $L_{yz}=(L_y^2+L_z^2)^{1/2}$, to be less or equal than the drag. In the case of an isolated oblate in Stokes flow, $L_{yz}/D$ is at most $0.12$ at $45^\circ$, while in the suspensions its median rises from $0.13$ at $\phi=0.2\%$ to $0.26$ at $15\%$, without dependence on $\Rey$. Interestingly, the tail of the distribution grows faster and at $\phi=15\%$, $41\%$ of the particles experience a lift larger than $0.3$ times their drag and $11\%$ larger than half of it. This is because with crowding, the distance between neighboring particles reduces and randomly oriented neighbors add large transverse forces that increase the lift of each particle. This agrees with the order of magnitude found by Sanjeevi\&Padding for spherocylinders at $\Rey=0.1$, as does the increase in the probability of large ratios with crowding \cite{sanjeevi2020}. If mean ratio found here grows with $\phi$, following the drag anisotropy, for prolate ellipsoids at $10\le\Rey\le200$ Cao\&Tafti found the mean lift to drag ratio to decrease with $\phi$, with the drag growing faster than the lift \cite{cao2022}.

Finally, a lift component $L_y$ measured to be less than $0.8\%$ of the mean drag, with no systematic sign and independent of orientation was detected from the particles in suspension. Compared with the work of De Souza et al.\ for an oblate spheroid in contact with a wall, the $L_y$ load of the present study lacks their $\sin^2\theta$ dependency that would be typical of a shear-induced lift, in agreement with the exclusion of the particles close to the free slip planes.

\subsubsection{Torque}\label{sec:res_torque}
Studies on fixed assemblies in uniform flows discussed in Section \ref{sec:intro} have shown that only the inertial pitching torque surviving on average \cite{sanjeevi2020}. However, in a shear flow, the viscous Jeffery torque is present at any $\Rey$ and dominates over the range investigated here. Accordingly, the torque acting on each particle in suspension was expressed as the sum of two contributions: the Jeffery torque $\mathbf{T}_J$ of Equation ~\eqref{eq:jeffery} and the pitching torque of Equations ~\eqref{eq:pitching} and \eqref{eq:cp} based on the isolated oblate simulations of Section \ref{sec:isolated_oblate}, each multiplied by a factor accounting for the effects of crowding. The two contributions can be readily distinguished because they exhibit distinct dependencies on particle orientation, $\dot\gamma$ and $U$. The expression reads
\begin{equation}
\mathbf{T}=g_T\,\mathbf{T}_J+h_P\,\mathbf{T}_P ,
\label{eq:torque_closure}
\end{equation}
where $g_T$ and $h_P$ are determined independently at each volume fraction by least-squares fitting to the three torque components of all retained particles.

\begin{figure*}
  \centering
  \includegraphics[width=0.99\textwidth]{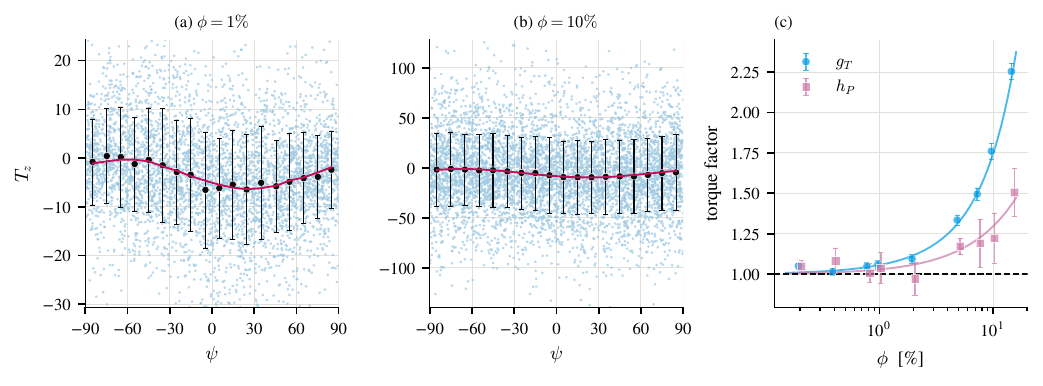}
  \caption{Normalized torque $T_z$ against $\psi$ at $\phi=1\%$ (a) and $10\%$ (b). In red, the closure of Eq.~\eqref{eq:torque_closure}. Blue dots are individual particles (at most $6\,000$ per panel). Black points are means over $10^\circ$ ranges, with standard deviation of the individual particles in each range. (c) Factors of the Jeffery and of the pitching torque. Squares are simulations with the standard error over the realizations. Curves are Eq.~\eqref{eq:gt}--\eqref{eq:hp}.}
  \label{fig:torque}
\end{figure*}

As for drag and lift, the torque acting on individual particles is strongly influenced by their neighbors, although this effect is more pronounced here. This results in fluctuations of $T_z$ within a given orientation range that are large compared to its mean, with the standard deviation reaching several times the mean value, as shown in Figure~\ref{fig:torque}(a,b). A similar behavior was reported by Cao et al. for prolate assemblies \cite{cao2022}.

The two crowding factors exhibit markedly different behaviours. The Jeffery factor, shown in Figure ~\ref{fig:torque} (c), remains close to unity up to $\phi=2\%$, despite the drag already growing by $20$--$60\%$ over this range, then rapidly increases with volume fraction up to $2.25$ at $\phi=15\%$. This suggests that crowding affects rotational resistance less strongly and at higher volume fractions than translational resistance and leads to the attribution of this difference to the faster spatial decay of the disturbance generated by a torque, especially when compared with that generated by a force. Consistently, if used to scale the torque, the drag factor would systematically overestimate it at all volume fractions. The pitching factor $h_P$, in turn, is less accurately determined as the pitching torque can be appreciated only at the upper end of the investigated $\Rey$ range, remaining between $0.97$ and $1.22$ up to $\phi=10\%$, before reaching $1.50$ at $\phi=15\%$. The resulting fits are \begin{align}
g_T&=1+5.45\,\phi+19.6\,\phi^2, \label{eq:gt} \\
h_P&=1+2.95\,\phi \label{eq:hp}
\end{align}
with maximum deviations of $4\%$ and $9\%$, respectively. The inertial torque therefore appears to increase less strongly with crowding than the viscous contribution. However, the scatter in $h_P$ prevents a firm conclusion and the linear dependence adopted for $h_P$ is the simplest form compatible with the data.

Sanjeevi\&Padding observed that neighboring spherocylinders induce no torque about the symmetry axis of a particle \cite{sanjeevi2020}. In the present suspensions, this result holds for the mean axial torque, indeed this component is captured by the axial component of the Jeffery torque simply multiplied by a crowding factor, with no additional contribution. This factor, obtained by least-squares fitting to $\mathbf{T}\cdot\mathbf{e}$ alone, agrees with $g_T$ to within $5.5\%$ up to $\phi=5\%$. However, at higher volume fractions, it exceeds $g_T$ by $6$--$9\%$. It can be deduced that, as observed for drag, crowding affects one principal resistance more strongly than the others and in this case, it is the resistance to rotation about the symmetry axis. Using a single factor in Equation \eqref{eq:torque_closure} therefore underestimates the mean axial torque by up to $9\%$ at $\phi=15\%$ and introducing a separate factor for rotation about the symmetry axis would remove this discrepancy, at the cost of an additional fitting parameter.

At the level of individual particles however, the axial torque can be substantial. Its standard deviation about the closure prediction is between $2$ and $19$ times the mean axial Jeffery torque, and increases with $\phi$. It follows that although the mean axial torque contains no contribution beyond the scaled Jeffery torque, individual particles experience significant axial torques that tend to drive rotation about their symmetry axes.

\subsection{Closure accuracy and limits}\label{sec:res:closure}
The proposed closure provides a unified description of the mean force and torque acting on a particle, combining the exact Stokes contributions characterized by $f_\parallel$, $f_\perp$, $X$, $Y$ and $Z$, with corrections accounting for finite Reynolds number and particle crowding. These corrections are expressed through eleven fitted constants, listed in Table~\ref{tab:coeff}. The resulting force and torque expressions are given by Eqs.~\eqref{eq:drag_closure}--\eqref{eq:lift_closure_z} and Eqs.~\eqref{eq:pitching}--\eqref{eq:cp} and \eqref{eq:torque_closure}--\eqref{eq:hp}, respectively. By construction, the closure recovers the Stokes expressions in the joint limit $\phi\to0$ and $\Rey\to0$.

The accuracy of the closure is assessed in Table~\ref{tab:accuracy}, which reports the mean and maximum deviations from the $10^\circ$-binned simulation means, following the approach of earlier correlation studies \cite{sanjeevi2018,sanjeevi2020}. Since lift and torque change sign with particle orientation, their deviations are normalized by the largest mean amplitude across the volume fractions, rather than by the local mean. The drag is reproduced to within $1.7\%$ for $\phi\ge2\%$ and $3.3\%$ at lower volume fractions. The lift, for which no fitting constant was introduced, is reproduced with mean deviations of $2$--$8\%$ of its amplitude, while the corresponding deviations for $T_z$ are $5$--$9\%$. The predictions are less accurate for $T_x$ and $T_y$, whose mean deviations range from $10$ to $26\%$, with maximum deviations of the order of their amplitudes. This reflects the small means of these components compared with the fluctuations induced by the neighboring particles, so that with the present number of oblates in suspension, the uncertainty of each $10^\circ$ mean is itself $13$--$45\%$ of their amplitude, against $7$--$10\%$ for $T_z$. It follows that the deviations of the closure would decrease with an even larger number of particles. De Souza et al. reported rolling and yawing torques, corresponding to $T_x$ and $T_y$ of the present study, reaching $0.3$ and $0.62$ of the pitching torque respectively and concluded that these contributions should not be neglected \cite{desouza2025}. In the present suspensions by contrast, their mean values are small, although the inter particle fluctuations are comparable to or larger than the means.
\begin{table}[t]
\caption{Mean $|$ maximum deviation (in $\%$) between the closure and the means of the simulations over $10^\circ$ ranges of the governing angle, all $\Rey$ ranges together.}\label{tab:accuracy}
\small
\begin{tabular}{@{}rcccccc@{}}
\toprule
$\phi$ [\%] & $D$ & $L_y$ & $L_z$ & $T_x$ & $T_y$ & $T_z$ \\
\midrule
$0.2$ & $1.6|2.6$ & $8|22$ & $5|15$ & $11|39$ & $10|33$ & $5|17$ \\
$0.4$ & $0.9|1.9$ & $7|20$ & $6|20$ & $16|45$ & $14|35$ & $5|11$ \\
$0.8$ & $0.9|2.5$ & $5|16$ & $5|13$ & $14|70$ & $13|38$ & $7|18$ \\
$1$   & $1.2|3.3$ & $7|16$ & $5|14$ & $22|114$ & $19|58$ & $9|29$ \\
$2$   & $0.3|0.9$ & $5|16$ & $6|31$ & $22|59$ & $16|41$ & $7|15$ \\
$5$   & $0.6|1.4$ & $3|6$ & $3|12$ & $22|90$ & $21|84$ & $9|26$ \\
$7.5$ & $0.6|1.7$ & $3|7$ & $4|12$ & $13|94$ & $17|53$ & $7|19$ \\
$10$  & $0.3|0.5$ & $2|7$ & $3|8$ & $26|121$ & $17|57$ & $6|17$ \\
$15$  & $0.3|0.7$ & $2|4$ & $2|11$ & $25|89$ & $26|107$ & $7|17$ \\
\bottomrule
\end{tabular}
\end{table}

The applicability of the closure is limited by the conditions under which it was developed. It describes fixed oblate particles with a single aspect ratio $\lambda=0.25$, in isotropic random suspensions with $0.2\%\le\phi\le15\%$ and $0.1\le\Rey\le1$ and away from boundaries. Its predictions represent the mean load on a particle of given orientation, not the load on an individual particle. The latter can depart from the predicted mean by about $20\%$ for the drag and by more than the mean itself for the torque. The recovery of the Stokes loads as $\phi\to0$ is also restricted to the limit $\Rey\to0$. At finite $\Rey$, an isolated particle experiences an Oseen correction to its drag \cite{brenner1961}, which is screened by neighboring particles in the suspensions considered here (Section~\ref{sec:res_drag}). Below $\phi=0.2\%$, this screening weakens, and the drag approaches the Oseen-corrected value for an isolated particle rather than the Stokes value. The closure should therefore not be extrapolated to more dilute suspensions at finite $\Rey$. Finally, since the closure has been developed on fixed assemblies, when modeling rotating particles its Stokes contribution can be extended by expressing the fluid rotation relative to the particle, making sure to correct the inertial pitching torque to account for this angular velocity difference.

\section{The proposed closure}\label{sec:closure_summary}
At last, the closure is collected here in the form in which it is meant to be implemented to perform CFD--DEM simulations. The loads are given in the frame of the shear flow, with $x$ the flow, $y$ the gradient and $z$ the vorticity direction for a particle of orientation $\mathbf{e}$ (Eq.~\eqref{eq:orientation}, $e_x\ge0$), in a suspension of local solid volume fraction $\phi$. The inputs are the slip $\mathbf{u}_s$ between the fluid velocity at the particle center and the particle velocity of magnitude $U$, the local shear rate $\dot\gamma$ and the slip Reynolds number $\Rey=U\deq/\nu$.

\paragraph{Force.} The drag, along the slip, is that of an isolated oblate particle with the resistances along and across the symmetry axis scaled by the crowding factors $g_\parallel$ and $g_\perp$:
\begin{equation}
F_D=3\pi\mu\deq U\left[f_\perp g_\perp+\left(f_\parallel g_\parallel-f_\perp g_\perp\right)e_x^2\right],
\label{eq:force_summary}
\end{equation}
\begin{equation}
g_k=1+a_k\,\phi^{1/2}+b_k\,\phi\ln\phi+c_k\,\phi,\qquad k=\perp,\parallel.
\end{equation}
The lift, in the gradient and vorticity directions, follows from the same factors with no separate lift coefficient:
\begin{align}
F_{L,y}&=3\pi\mu\deq U\left(f_\parallel g_\parallel-f_\perp g_\perp\right)e_xe_y, \nonumber\\
F_{L,z}&=3\pi\mu\deq U\left(f_\parallel g_\parallel-f_\perp g_\perp\right)e_xe_z.
\label{eq:lift_summary}
\end{align}

\paragraph{Torque.} The torque is the sum of the viscous Jeffery torque and of the inertial pitching torque, each scaled by its own crowding factor:
\begin{equation}
\mathbf{T}=g_T\,\mu\dot\gamma\deq^3\,\tilde{\mathbf{T}}_J(\mathbf{e})+h_P\,c_P\,\rho\,U^2\deq^3\,e_x\left(\mathbf{e}\times\hat{\mathbf{x}}\right),
\label{eq:torque_summary}
\end{equation}
\begin{align}
g_T&=1+5.45\,\phi+19.6\,\phi^2, \nonumber\\
h_P&=1+2.95\,\phi, \\
c_P&=1.049\left(1-0.157\,\Rey\right), \nonumber
\end{align}
where $\tilde{\mathbf{T}}_J$ has the components $T_x$, $T_y$ and $T_z$ of Eq.~\eqref{eq:jeffery}.

\paragraph{Parameters and validity.} The exact Stokes values for $\lambda=0.25$, the eleven fitted constants and the range of validity are listed in Table~\ref{tab:coeff}. Within this range the closure gives the mean load on a particle of given orientation: the drag within $1.7\%$ ($3.3\%$ for $\phi<2\%$), the lift within $2$--$8\%$ of its amplitude and the torque about the vorticity axis within $5$--$9\%$, while the out-of-plane torques $T_x$ and $T_y$ are only captured qualitatively. The load on an individual particle departs from the mean by about $20\%$ for the drag and by more than the mean for the torque, depending on its neighbors. For $\phi\to0$ and $\Rey\to0$ the closure reduces to the exact Stokes loads, but it should not be extrapolated below $\phi=0.2\%$ at finite $\Rey$, where the drag approaches the Oseen-corrected value of the isolated particle.

\begin{table}[t]
\caption{Parameters and range of validity of the closure, Eqs.~\eqref{eq:force_summary}--\eqref{eq:torque_summary}.}\label{tab:coeff}
\small
\begin{tabular}{@{}lccc@{}}
\toprule
\multicolumn{4}{@{}l}{Exact Stokes values, $\lambda=0.25$ \cite{happel1983,jeffery1922}} \\
\midrule
Translational & \multicolumn{3}{l}{$f_\parallel=1.3768$, \quad $f_\perp=1.0831$} \\
Rotational & \multicolumn{3}{l}{$X=7.0671$, \quad $Y=5.7903$} \\
Strain coupling & \multicolumn{3}{l}{$Z=-5.1091$} \\
\midrule
\multicolumn{4}{@{}l}{Fitted constants} \\
\midrule
 & $a_k$ & $b_k$ & $c_k$ \\
$g_\perp$ & $3.910$ & $5.324$ & $19.68$ \\
$g_\parallel$ & $6.226$ & $10.65$ & $31.80$ \\
$g_T$ & \multicolumn{3}{l}{$5.45$, \quad $19.6$} \\
$h_P$ & \multicolumn{3}{l}{$2.95$} \\
$c_P$ & \multicolumn{3}{l}{$1.049$, \quad $0.157$} \\
\midrule
\multicolumn{4}{@{}l}{Range of validity} \\
\midrule
Aspect ratio & \multicolumn{3}{l}{$\lambda=0.25$} \\
Volume fraction & \multicolumn{3}{l}{$0.2\%\le\phi\le15\%$} \\
Slip Reynolds & \multicolumn{3}{l}{$0.1\le\Rey\le1$} \\
Orientations & \multicolumn{3}{l}{isotropic random suspension} \\
Boundaries & \multicolumn{3}{l}{farther than $8\,\deq$} \\
Particles & \multicolumn{3}{l}{fixed (no particle rotation)} \\
\bottomrule
\end{tabular}
\end{table}

\section{Conclusion}
This work developed a closure for the drag, lift and torque acting on oblate particles in suspension under linear shear flow by means of particle-resolved lattice Boltzmann simulations. Fixed, isotropic random suspensions of oblate ellipsoids of aspect ratio $\lambda=0.25$, resembling blood platelets, were simulated at solid volume fractions $0.2\%\le\phi\le15\%$ over the range $0.1\le\Rey\le1$ and complemented by simulations of an isolated oblate.

The simulations of the isolated oblate provided the dilute limit and allowed the effects of slip and shear to be separated, a distinction that near-wall studies cannot draw. The transverse force was found to follow closely the difference between face-on and edge-on drag, with deviations remaining below 0.8\% for the out-of-plane configuration and increasing to 7.9\% at Re = 1 for the in-plane configuration. Concerning the torque, besides the viscous Jeffery torque, a pitching torque identified as the inertial torque was observed to grow with the square of the slip, with a coefficient tending to the theory of Dabade et al.\ \cite{dabade2015} for $\Rey\to0$.

In suspension, the loads on individual particles were found to scatter widely around their means as a consequence of the diversity of their neighborhoods. On average however, the drag retained the $\cos^2\theta$ dependence on the incidence angle of the isolated oblate at every volume fraction, thus extending the persistence of this scaling to crowded suspensions in linear shear flow. Moreover, both drag factors grew monotonically with $\phi$ already from the most dilute suspensions, consistently with the $\phi^{1/2}$ screening of a fixed bed, something that any spherical voidage function such as the one by Di Felice would under-predict. Their dependence on $\Rey$ was found to be limited to a few percent and confined to the most dilute suspensions, where the screening by the neighbors is weakest. The lift was proven to follow directly from the drag anisotropy in both the gradient and the vorticity directions, without any fitted lift constant and away from the boundaries, no shear-induced lift was detected. Finally, the viscous Jeffery torque was affected by crowding less strongly and at higher volume fractions than the drag, while the inertial pitching torque appeared to increase less strongly with crowding. Instead, similarly to the drag, crowding affected the resistance to rotation about the symmetry axis more strongly than the others.

Since the closure was derived for fixed, isotropically oriented oblates of a single aspect ratio away from boundaries, future investigations will involve the extension of the closure to other aspect ratios and above all will focus on suspensions of mixed particle shapes, as platelets in blood are surrounded by red blood cells. Nevertheless, to the best of the authors' knowledge, the present work provides the first closure for drag, lift and torque of oblate particles in suspension under linear shear flow, ready to be implemented in CFD-DEM simulations of large-scale platelet-laden flows, such as those in stenosed vessels.

\bibliographystyle{model1-num-names}
\bibliography{cas-refs}

\end{document}